\documentclass[journal]{IEEEtran}
\usepackage{cite}
\usepackage{amsmath,amssymb,amsfonts}
\usepackage{algorithm}
\usepackage{algorithmic}
\usepackage{graphicx}                
\usepackage{algorithm,algorithmic}
\usepackage{textcomp}
\usepackage[acronym]{glossaries}       
\usepackage{multirow}                  
\usepackage{array}
\usepackage{lipsum}                 
\usepackage{stfloats}                 
\usepackage{caption}
\usepackage{subcaption}  

\usepackage{booktabs}                  
\usepackage{verbatim}                

\usepackage{enumitem}                  
\usepackage{graphicx}                  
\usepackage{threeparttable}            
\usepackage[table,xcdraw]{xcolor}      
\makeatletter
\let\NAT@parse\undefined
\makeatother
\usepackage{url}                       
\usepackage{hyperref}                  

\def\BibTeX{{\rm B\kern-.05em{\sc i\kern-.025em b}\kern-.08em
    T\kern-.1667em\lower.7ex\hbox{E}\kern-.125emX}}
\makeglossaries
    \newacronym{ssvep}{SSVEP}{steady-state visual evoked potential}
    \newacronym{bci}{BCI}{brain-computer interface}
    \newacronym{snr}{SNR}{signal-to-noise ratio}
    \newacronym{cca}{CCA}{canonical correlation analysis}
    \newacronym{mcca}{MCCA}{multiway CCA}
    \newacronym{l1mcca}{L1-MCCA}{L1-regularized multiway CCA}
    \newacronym{msetcca}{MsetCCA}{multiset CCA}
    \newacronym{itcca}{ITCCA}{individual template-based CCA}
    \newacronym{pcca}{P-CCA}{phased-constrained CCA}
    \newacronym{trca}{TRCA}{task-related component analysis}
    \newacronym{eeg}{EEG}{electroencephalogram}
    \newacronym{corca}{CORCA}{correlated component analysis}
    \newacronym{psda}{PSDA}{power spectral density analysis}
    \newacronym{ttcca}{tt-CCA}{transfer template-based canonical correlation analysis}
    \newacronym{cssft}{CSSFT}{cross-subject spatial filter transfer}
    \newacronym{lst}{LST}{least-squares transformation}
    \newacronym{stcca}{stCCA}{subject transfer based CCA}
    \newacronym{iismc}{IISMC}{inter and intra-subject maximal correlation}
    \newacronym{transrca}{TransRCA}{transfer-related component analysis}
    \newacronym{iiscca}{IISCCA}{intra- and inter-subject CCA}
    \newacronym{ass}{ASS}{accuracy-based subject selection}
    \newacronym{ste}{STE}{subject transferability estimation}
    \newacronym{jfpm}{JFPM}{joint frequency and phase modulation}
    \newacronym{itrca}{iTRCA}{instance-based task-related component analysis}
    \newacronym{trc}{TRC}{task-related component}
    \newacronym{ssitrca}{SS-iTRCA}{subject selection-based iTRCA}
    \newacronym{itr}{ITR}{information transfer rate}
    \newacronym{dgtf}{DGTF}{domain generalization-based transfer learning}
    \newacronym{sem}{SEM}{stand error of mean}
    \newacronym{tsne}{t-SNE}{t-distributed stochastic neighbor embedding}
    \newacronym{chi}{CHI}{Calinski-Harabasz Index}
    \newacronym{tdcca}{TDCCA}{training data-driven CCA}
    \newacronym{a2o}{A2O}{all-to-one}
    \newacronym{o2o}{O2O}{one-to-one}
    \newacronym{loso}{LOSO}{leave-one-subject-out}
    \newacronym{lobo}{LOBO}{leave-one-block-out}
  
\begin{document}

\title{
    Instance-Based Transfer Learning with Similarity-Aware Subject Selection for Cross-Subject SSVEP-Based BCIs
        }

\author{
	Ziwen Wang$^{1}$, 
	Yue Zhang$^{2}$, 
	Zhiqiang Zhang$^{3}$, 
	Sheng Quan Xie$^{3}$, 
	Alexander Lanzon$^{1}$, 
	William P. Heath$^{4}$, 
	and Zhenhong Li$^{1}$ 
	\vspace{1mm} \\ 
	$^{1}$University of Manchester\quad
	$^{2}$University of Bath\quad
	$^{3}$University of Leeds\quad
	$^{4}$Bangor University

\thanks{IEEE Journal of Biomedical and Health Informatics.}
\thanks{DOI: \href{https://ieeexplore.ieee.org/document/11028588}{10.1109/JBHI.2025.3577813}}
}

\maketitle


\begin{abstract}
 Steady-state visual evoked potential (SSVEP)-based brain-computer interfaces (BCIs) can achieve high recognition accuracy with sufficient training data. Transfer learning presents a promising solution to alleviate data requirements for the target subject by leveraging data from source subjects; however, effectively addressing individual variability among both target and source subjects remains a challenge. This paper proposes a novel transfer learning framework, termed instance-based task-related component analysis (iTRCA), which leverages knowledge from source subjects while considering their individual contributions. iTRCA extracts two types of features: (1) the subject-general feature, capturing shared information between source and target subjects in a common latent space, and (2) the subject-specific feature, preserving the unique characteristics of the target subject. To mitigate negative transfer, we further design an enhanced framework, subject selection-based iTRCA (SS-iTRCA), which integrates a similarity-based subject selection strategy to identify appropriate source subjects for transfer based on their task-related components (TRCs). Comparative evaluations on the Benchmark, BETA, and a self-collected dataset demonstrate the effectiveness of the proposed iTRCA and SS-iTRCA frameworks. This study provides a potential solution for developing high-performance SSVEP-based BCIs with reduced target subject data.
\end{abstract}


\section{Introduction}
\label{sec:introduction}

\IEEEPARstart{S}{svep}--based \glspl{bci} facilitate direct communication between the brain and external devices \cite{wolpaw_2002_BCI_definition}. They have been widely used in robotics \cite{shao_2020_SSVEP_robot,chen_2019_ssvep_robotarm,chiuzbaian_2019_ssvep_drone}, rehabilitation \cite{vinoj_2019_ssvep_exoskeleton,zhao_2016_ssvep_rehabilitation,na_2021_ssvep_wheelchair,muller_2007_SSVEP_prosthesis}, and entertainment \cite{wang_2019_ssvep_game} due to their high \gls{itr}, high \gls{snr}, noninvasiveness, and minimal training needs\cite{bin_2009_VEP,li_2024_SUTL}. 

Precise decoding of \gls{ssvep} from \gls{eeg} signals is crucial for \gls{ssvep}-based \glspl{bci}. Existing methods can be broadly categorized into training-free (e.g., \cite{wang_2006_PSDA,lin_2006_CCA}) and training-based approaches (e.g., \cite{zhang_2011_multiwayCCA,bin_2011_ITCCA,zhang_2014_multisetCCA,zhang_2013_L1_multiwayCCA,zhang_2018_CORCA,nakanishi_2017_TRCA}). Compared to training-free approaches, training-based approaches usually achieve higher accuracy, provided that sufficient training data is collected during the calibration process, which is often time-consuming and labor-intensive.

Cross-subject \color{black} transfer learning offers a potential solution to alleviate data requirements for the target subject (i.e., the new subject whose SSVEP signals need to be recognized) by transferring knowledge from source subjects (i.e., subjects whose data have been previously collected). 
 Existing transfer learning approaches fall into two categories: one-to-one and all-to-one approaches. The one-to-one approach transfers knowledge from each source subject to the target subject individually and integrates the results using a grand average to form the final recognition criterion. Examples include \gls{iismc}\cite{wang_2021_IISMC},  \gls{dgtf}\cite{huang_2023_domain_gen_TF} and one-to-one cross-subject TRCA in \cite{liu_2019_cross_subject_TRCA}. The all-to-one approach, by contrast, constructs a unified representation of all source subjects through grand averaging, then transfers knowledge based on this integrated representation.   Examples include \gls{transrca}\cite{lan_2023_TransRCA}, all-to-one cross-subject TRCA in \cite{liu_2019_cross_subject_TRCA}, \gls{iiscca}\cite{wei_2023_IISCCA} and \gls{cssft}\cite{yan_2022_CSSFT}. However, the grand average in  one-to-one and all-to-one approaches both implicitly assume equal contributions from all source subjects during transfer, overlooking the substantial variability across source subjects.
 To address this, the pioneering algorithm in \cite{zhang_2023_yueTF} assigns contribution scores to source subjects in the final recognition criterion based on normalized correlation coefficients.  However, these coefficients are assigned after knowledge transfer, and how to account for the varying contributions of source subjects during knowledge transfer remains an open problem.

Furthermore, not all source subjects contribute positively to the target subject's \gls{ssvep} recognition \cite{pan_2009_survey_TF,rosenstein_2005_transfer_or_not}. When the knowledge learned from source subjects have a detrimental effect on the target subject (e.g., reducing recognition accuracy), the transfer is regarded as negative transfer \cite{weiss_2016_survey_of_TF}. To mitigate this phenomenon, pioneering work has begun to develop subject selection strategies prior to transfer, which generally follow a two-step approach: (1) evaluate each source subject's transferability based on recognition accuracy; (2) select several top-performing subjects to transfer, e.g. \gls{cssft} \cite{yan_2022_CSSFT}, \gls{ass} \cite{wei_2023_IISCCA}, and \gls{ste} \cite{li_2024_SUTL}. However, these accuracy-based selection strategies require a complete recognition process for each source subject prior to transfer learning, imposing computational load and extended processing time \cite{wei_2023_IISCCA}.

In this paper, we propose a new transfer learning framework, \gls{itrca}, for cross-subject \gls{ssvep} recognition. Motivated by the private-share component analysis in neural network design \cite{bousmalis_2016_domain_seperation_network}, \gls{itrca} innovatively employs a dual-feature approach that combines both the general characteristics across subjects and the unique characteristics of the target subject.

\begin{enumerate}
    \item Subject-general feature: this feature captures the general characteristics of \gls{ssvep} response across source and target subjects through transfer learning. During the transfer learning, we innovatively formulate each source subject's \gls{ssvep} response as an instance and construct a common latent space between source instances and the target subject, considering different contributions of source instances. In this space, the source instances and the target subject's SSVEP response are maximally correlated.
    \item Subject-specific feature: this feature captures the individual characteristics of the target subject's \gls{ssvep} response based on \gls{trca}.
\end{enumerate}
 To avoid negative transfer, we further propose a similarity-based subject selection strategy to filter out source instances that are dissimilar to the target. Different from the existing subject selection strategies, the proposed strategy quantifies the similarity based on subjects' \glspl{trc},  thereby eliminating the need for a complete recognition process for each source subject. Comparative studies are carried out on three datasets: Benchmark, BETA, and a self-collected dataset, and evaluated against state-of-the-art methods \gls{transrca} \cite{lan_2023_TransRCA}, \gls{dgtf} \cite{huang_2023_domain_gen_TF}, and \gls{cssft} \cite{yan_2022_CSSFT}. The results demonstrate the superior recognition performance of \gls{itrca} compared to baseline methods, and further enhanced through the integration of subject selection strategy. Additionally, our proposed similarity-based subject selection strategy substantially reduces computational cost compared to the accuracy-based method.

\renewcommand{\arraystretch}{1.4}  
\begin{table*}[!t]                 
                                
 \caption{Datasets Description}
 \label{tab:datasets comparison}
 \centering
 
    \begin{tabular}{c|l|l|l}
        \toprule
        \textbf{Dataset}  &\textbf{Benchmark}  &\textbf{BETA}  &\textbf{Self-collected } \\ 
        \hline
        \textbf{Subjects} 
            &35 healthy subjects (17F / 18M)    
            &70 healthy subjects (28F / 42M) 
            &11 healthy subjects (5F / 6M) \\ 
        \hline
        \multirow{2}{*}{\textbf{Stimuli}}
            & 40 stimuli    & 40 stimuli   & 12 stimuli \\
            & $f_i = 8+0.2(i-1)$ Hz, $i = 1,2,\ldots,40.$ 
            & $f_i = 8+0.2(i-1)$ Hz, $i = 1,2,\ldots,40.$ 
            & $f_i = 9.25+0.5(i-1)$ Hz, $i = 1,2,\ldots,12.$ \\
        \hline
        \textbf{Trials}
            & 40 trials / 6 blocks  
            & 40 trials / 4 blocks
            & 12 trials / 5 blocks \\ 
        \hline
        \textbf{Time}
            & 0.5 s cue + 5 s flickering + 0.5 s rest
            & 0.5 s cue + 2 s or 3 s flickering + 0.5 s rest
            & 0.5 s cue + 5 s flickering + 0.5 s rest \\  
        \bottomrule

    \end{tabular}
\end{table*}

\section{Methods}
\label{sec:methods}

\subsection{Dataset}
\label{subsec:datasets}

In this study, three datasets were utilized: two public datasets (Benchmark \cite{wang_2016_benchmark} and BETA \cite{liu_2020_beta}) generated by Tsinghua University, and a self-collected dataset generated by the University of Leeds. Essential details about three datasets are summarized in Table \ref{tab:datasets comparison}. 

\begin{figure}[!t]
    \centering
    \includegraphics[width=1\linewidth]{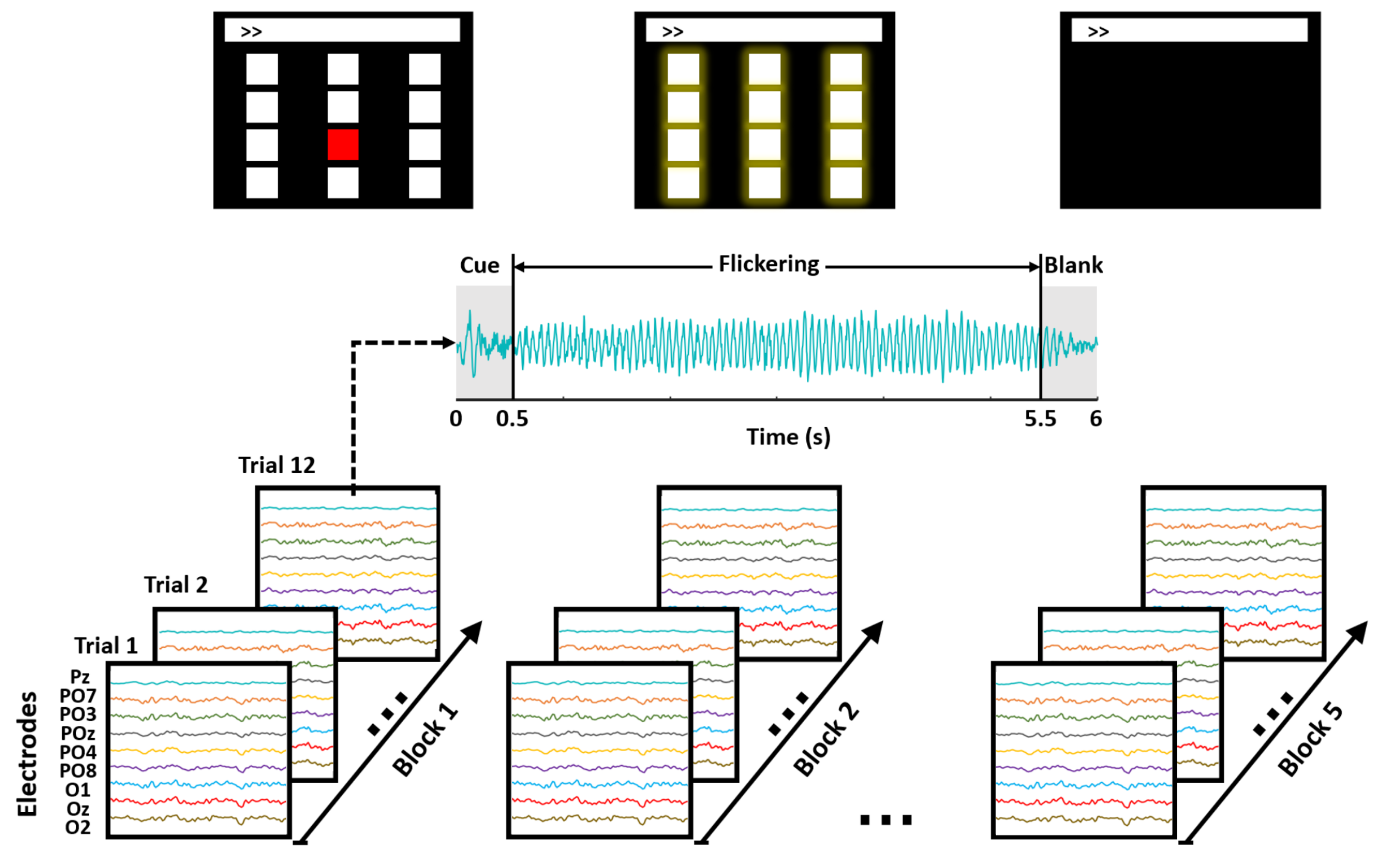}
    \vspace{0.05cm}
    \caption{The experiment paradigm of the self-collected dataset.}
    \label{fig:experiment}
\end{figure}

Fig. \ref{fig:experiment} illustrates the experiment paradigm of the self-collected dataset. The interaction interface was composed of 12 flickering stimuli arranged in a 4 $\times$ 3 matrix and displayed on the 23.6-inch LCD monitor with a resolution of 1920 $\times$ 1080 pixels and a refresh rate of 60 Hz. The stimuli were encoded using \gls{jfpm} approach \cite{zhang_2014_JFPM}, where frequencies varied from 9.25 Hz to 14.75 Hz in 0.5 Hz increments and phase shifts cycled from 0$\pi$ to 1.5$\pi$ in 0.5$\pi$ increments. The experiment consisted of 5 blocks, each containing 12 trials corresponding to different stimuli. Each trial began with a 0.5 s red dot cue indicating the target flicker, followed by 5 s of gaze fixation on the target, and concluded with 0.5 s of rest before the next trial. \Gls{eeg} signals were recorded at 256 Hz using g.USBamp (g.tec medical engineering) and then band-pass filtered between 8 and 40 Hz using the IIR filter of Chebyshev type I. 
The experiment was approved by the Research Ethics Committee of the University of Leeds (reference number MEEC 21-024), and all participants signed the informed consent.
To ensure consistency in data processing, the signals from Benchmark and BETA datasets were band-pass filtered between 8 Hz and 88 Hz using the IIR filter of Chebyshev type I.

Across all three datasets, nine electrodes located in parietal and occipital areas (Pz, PO7, PO3, POz, PO4, PO8, O1, Oz, and O2) were employed for three datasets. \Gls{eeg} signals were segmented within the window [0.64, 0.64 + $d$] s, accounting for the 0.5 s gaze shifting and 0.14 s latency delay, where $d$ denotes the data length used for \gls{ssvep} recognition.

\subsection{Notations}

For each subject, the multi-channel \gls{eeg} signals are denoted by a 4-th order tensor. For source subjects, the available training data are represented by
\begin{equation}
    \{\boldsymbol{\mathcal{X}}^{S_n}\in \mathbb{R}^{N_f\times N_c\times N_s \times N_{sb}}\}_{n=1}^{N_{src}} \nonumber
\end{equation} 
with entries $\boldsymbol{\mathcal{X}}^{S_n}(i,p,q,h)\in \mathbb{R}$, where $n$ and $N_{src}$ denote the index and total number of source subjects, respectively. For the target subject, the available training data are represented by 
\begin{equation}
    \boldsymbol{\mathcal{X}}^{T_a}\in \mathbb{R}^{N_f\times N_c\times N_s \times N_{tb}} \nonumber
\end{equation} 
with entries $\boldsymbol{\mathcal{X}}^{T_{a}}(i,p,q,j)\in \mathbb{R}$. The indices and dimensions of tensors are defined as follows:
\begin{itemize}[label=\scalebox{0.7}{$\bullet$}]
    \item $i$ and $N_f$: The index and total number of stimuli.
    \item $p$ and $N_c$: The index and total number of channels.
    \item $q$ and $N_s$: The index and total number of samples.
    \item $h$ and $N_{sb}$: The index and total number of source blocks.
    \item $j$ and $N_{tb}$: The index and total number of target blocks.
\end{itemize}
For simplicity, we will use the source subject's tensor $\boldsymbol{\mathcal{X}}^{S_n}$ as an example to introduce the definition of notation. For the $n$-th source subject and $i$-th stimulus, a single trial in the $h$-th block is denoted as the 2-D matrix $\mathbf X_{i,h}^{S_n}=\boldsymbol{\mathcal{X}}^{S_n}(i,:,:,h)\in\mathbb{R}^{N_c\times N_s}$, trials from all blocks are denoted as $\mathbf X_i^{S_n}=\boldsymbol{\mathcal{X}}^{S_n}(i,:,:,:)\in\mathbb{R}^{N_c\times N_s \times N_{sb}}$, the individual template is the grand average across all blocks, i.e., $\bar{\mathbf X}_i^{S_n} = 1/N_{sb}\sum_{h=1}^{N_{sb}}\mathbf{X}_{i,h}^{S_n} \in \mathbb{R}^{N_c\times N_s}$, and their ground truth label is the visual stimulus frequency $f_i$. These rules of notation definition also apply to the target subject's tensor $\boldsymbol{\mathcal{X}}^{T_a}$.

In this study, we consider $N_f$ visual stimuli at frequencies $f_1, f_2,\ldots, f_{N_f}$. Given a test trial $\mathbf{\Tilde{X} }\in \mathbb{R}^{N_c\times N_s}$ from the target subject, we aim to recognize the corresponding stimulus frequency $\hat f\in \{f_1, f_2, \ldots, f_{N_f}\}$, using both training data from the target subject $\boldsymbol{\mathcal{X}}^{T_a}$ and source subjects $\{\boldsymbol{\mathcal{X}}^{S_n}\}_{n=1}^{N_{src}}$.

\subsection{\gls{itrca}}

\begin{figure*}
    \centering
    \includegraphics[width=0.93\linewidth]{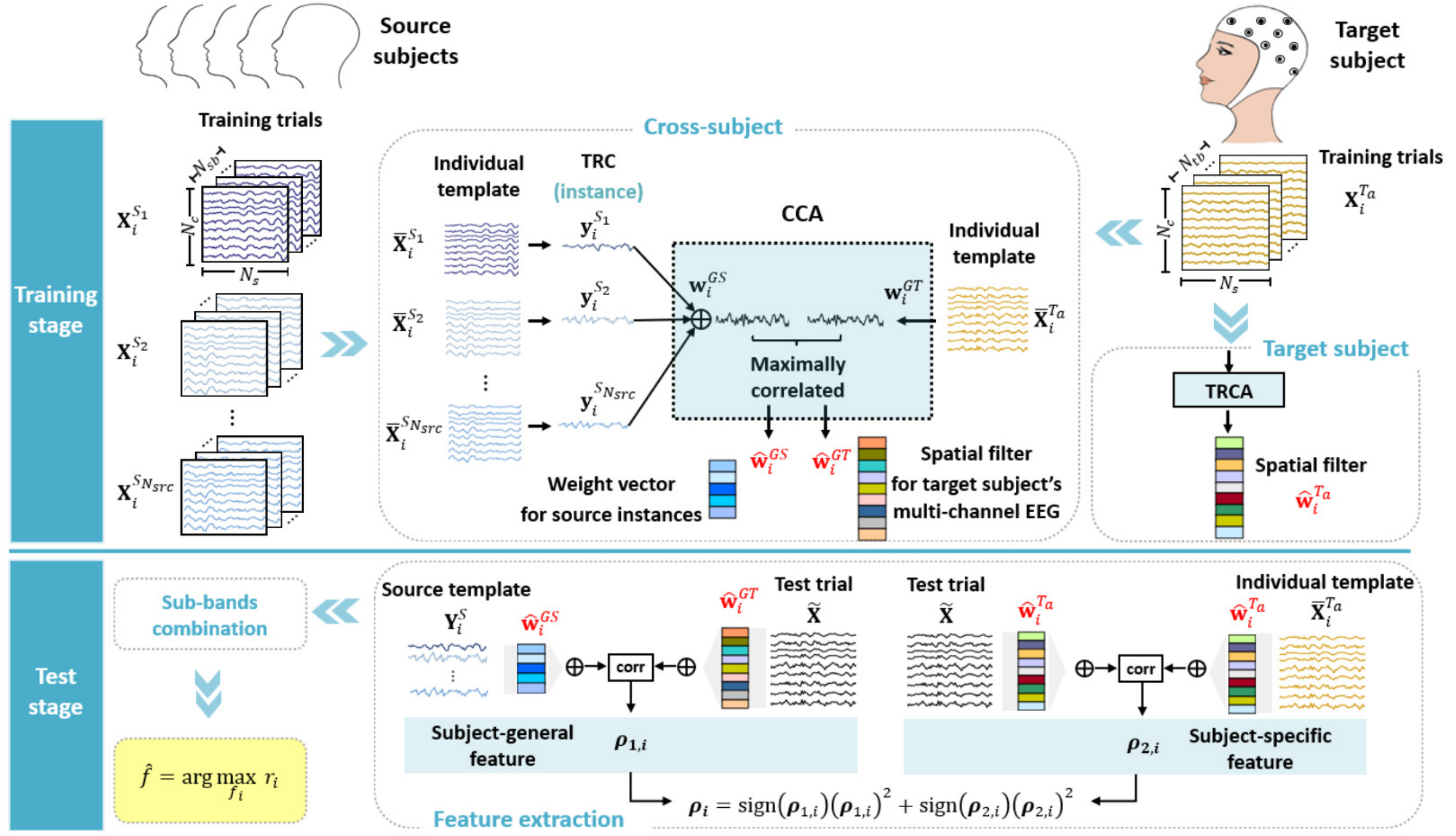}
    \caption{Diagram of the proposed \gls{itrca} framework. For the $i$-th stimulus, the shared information across subjects is captured in a common latent space (the blue block) where source instances and the target subject's individual template are maximally correlated through weight vector $\hat{\mathbf{w}}_i^{GS}$ and the spatial filter $\hat{\mathbf{w}}_i^{GT}$. Meanwhile, target subject's individual information is captured by the spatial filter $\hat{\mathbf{w}}_i^{T_a}$ based on \gls{trca}. In the test stage, the feature $\rho_i$ is composed of the subject-general feature $\rho_{1,i}$ and the subject-specific feature $\rho_{2,i}$, and then combined across all sub-bands to formulate the feature $r_i$ used for \gls{ssvep} recognition.}
    \label{fig:iTRCA}
\end{figure*}

\gls{itrca} is a transfer learning framework for \gls{ssvep} cross-subject recognition, incorporating both subject-general and subject-specific features. Inspired by the concept of instance-based transfer learning \cite{wang_2023_TF_WangJindong}, \gls{itrca} computes the \glspl{trc} of source subjects as source instances and then projects them into a latent space, where these source instances and the target subject's template exhibit maximal correlation. This latent space captures the underlying characteristics of SSVEP responses shared between source and target subjects, while accounting for the varying contributions of individual source subjects. Meanwhile, \gls{itrca} captures target subject's individual characteristics using \gls{trca}. The overall framework is illustrated in Fig. \ref{fig:iTRCA} and includes following steps.

\subsubsection{Compute the source instances} 
\gls{trc}, which is the single-channel signal spatially filtered from multi-channel \gls{eeg} signal based on \gls{trca} \cite{nakanishi_2017_TRCA}, is chosen as the instance of each source subject due to its low-dimensional nature and demonstrated effectiveness in recovering SSVEP characteristics \cite{tanaka_2020_gtrca, lan_2023_TransRCA}.

For the $n$-th source subject and $i$-th stimulus, the spatial filter $\mathbf{w}_i^{S_n}\in\mathbb{R}^{N_c}$ is invariant across different trials. First, the spatially filtered inter-trial covariance across all combinations is summed together

\begin{align}
        &\sum_{\substack{h_1,h_2=1 \\ h_2\neq h_1}}^{N_{sb}} 
            \mathrm{Cov}\left( (\mathbf{w}_i^{S_n})^{\mathsf{T}}\mathbf{X}_{i,h_1}^{S_n},(\mathbf{w}_i^{S_n})^{\mathsf{T}}\mathbf{X}_{i,h_2}^{S_n} \right) \nonumber
     \\ = & \sum_{\substack{h_1,h_2=1 \\ h_2\neq h_1}}^{N_{sb}} 
            (\mathbf{w}_i^{S_n})^{\mathsf{T}}   
            \mathrm{Cov}(\mathbf{X}_{i,h_1}^{S_n},\mathbf{X}_{i,h_2}^{S_n}) 
            \mathbf{w}_i^{S_n} \nonumber
     \\ = & (\mathbf{w}_i^{S_n})^{\mathsf{T}}
            \mathbf{S}_i^{S_n} 
            \mathbf{w}_i^{S_n},
\label{eq:inter-trial covariance}
\end{align}
where 
\begin{equation}
\begin{aligned}
     \mathbf{S}_i^{S_n} = \sum_{\substack{h_1,h_2=1 \\ h_2\neq h_1}}^{N_{sb}}  
                          \mathrm{Cov}(\mathbf{X}_{i,h_1}^{S_n},\mathbf{X}_{i,h_2}^{S_n}).
\end{aligned}
\end{equation}
Then, the variance of the spatially filtered signal is constrained to 1
\begin{align}
               & \sum_{h=1}^{N_{sb}} 
                 \mathrm{Cov}\left( (\mathbf{w}_i^{S_n})^{\mathsf{T}}\mathbf{X}_{i,h}^{S_n},(\mathbf{w}_i^{S_n})^{\mathsf{T}}\mathbf{X}_{i,h}^{S_n} \right) \nonumber
           \\ = & \sum_{\substack{h=1}}^{N_{sb}} 
                (\mathbf{w}_i^{S_n})^{\mathsf{T}}   
                 \mathrm{Cov}(\mathbf{X}_{i,h}^{S_n},\mathbf{X}_{i,h}^{S_n}) 
                 \mathbf{w}_i^{S_n} \nonumber 
           \\ = & (\mathbf{w}_i^{S_n})^{\mathsf{T}}
                 \mathbf{Q}_i^{S_n} 
                 \mathbf{w}_i^{S_n}  \nonumber 
           \\ = & 1, 
\label{eq:recovered trial variance}
\end{align}
where 
\begin{equation}
\begin{aligned}
     \mathbf{Q}_i^{S_n} = \sum_{\substack{h=1}}^{N_{sb}}  
                          \mathrm{Cov}(\mathbf{X}_{i,h}^{S_n},\mathbf{X}_{i,h}^{S_n}).
\end{aligned}
\end{equation}
The spatial filter $\mathbf{w}_i^{S_n}$ is computed by maximizing (\ref{eq:inter-trial covariance}) under the constraint (\ref{eq:recovered trial variance}), which is in the form of Rayleigh Quotient \cite{horn_2012_matrixAnalysis(RayleighRitz)}
\begin{equation}
    \mathbf{\hat w}_i^{S_n} 
    = \mathop{\arg\max}\limits_{\mathbf{w}_i^{S_n}} 
      \frac{(\mathbf{w}_i^{S_n})^{\mathsf{T}}\mathbf{S}_i^{S_n}\mathbf{w}_i^{S_n}}
           {(\mathbf{w}_i^{S_n})^{\mathsf{T}}\mathbf{Q}_i^{S_n}\mathbf{w}_i^{S_n}}.
\label{eq:Rayleigh Quotient}
\end{equation}
Therefore, the optimal spatial filter can be solved as the eigenvector of matrix $(\mathbf{Q}_i^{S_n})^{-1}\mathbf{S}_i^{S_n}$ corresponding to the largest eigenvalue. For simplicity, the calculation of $\mathbf{\hat w}_i^{S_n} $  from (\ref{eq:inter-trial covariance}) to (\ref{eq:Rayleigh Quotient}) is defined as the following function
\begin{equation}
    \mathbf{\hat w}_i^{S_n} = \mathrm{TRCA}(\mathbf{X}_i^{S_n}),
\label{eq:TRCA}
\end{equation}
where the input $\mathbf{X}_i^{S_n}$ represents all training trials for the $n$-th source subject and $i$-th stimulus.
The single-channel \gls{trc} $\mathbf{y}_i^{S_n}$ is extracted from the individual template $\mathbf{\Bar{X}}_i^{S_n}$
\begin{equation}
    \mathbf{y}_i^{S_n} =  (\mathbf{\hat w}_i^{S_n})^{\mathsf{T}} \mathbf{\Bar{X}}_i^{S_n} \in \mathbb{R}^{1\times N_s}.
\end{equation}
The \glspl{trc} from all source subjects are vertically concatenated as the source template, 
\begin{equation}   
\label{eq:source instances}
\mathbf{Y}_i^S = \left[\begin{array}{c} \mathbf{y}_i^{S_1} \\
                                        \mathbf{y}_i^{S_2} \\
                                          \vdots \\
                                \;\;\;\;\mathbf{y}_i^{S_{N_{src}}} \\
                       \end{array}\right] 
\in \mathbb{R}^{N_{src}\times N_s},
\end{equation}
which is the unified representation of source subjects.

\subsubsection{Weight source instances} 
\label{subsection:Weight source instances}
 For the $i$-th stimulus, the source template $\mathbf{Y}_i^S$ and the target subject's individual template $\bar{\mathbf X}_i^{T_a} = 1/N_{tb}\sum_{j=1}^{N_{tb}}\mathbf{X}_{i,j}^{T_a} \in \mathbb{R}^{N_c\times N_s}$  are projected to a common latent space where they are maximally correlated by the source weight vector $\mathbf{\hat w}_i^{GS}\in \mathbb{R}^{N_{src}}$ and the target spatial filter $\mathbf{\hat w}_i^{GT}\in \mathbb{R}^{N_c}$ based on \gls{cca} \cite{mardia_2024_ccaBook} as follows: 

\begin{equation}
\begin{aligned}
    \label{eq:CCA}
    [\mathbf{\hat w}_i^{GS}, \mathbf{\hat w}_i^{GT}] 
    & = \mathop{\arg\max}\limits_{\mathbf{w}_i^{GS}, \mathbf{w}_i^{GT}}\;
        \mathrm{corr}((\mathbf{w}_i^{GS})^{\mathsf{T}}\mathbf{Y}_i^S, (\mathbf{w}_i^{GT})^{\mathsf{T}}\mathbf{\bar X}_i^{T_a}), \\
\end{aligned}
\end{equation}
where $\mathrm{corr}(\cdot,\cdot)$ is Pearson correlation.

\subsubsection{Feature extraction}

\gls{itrca} extracts two kinds of features: subject-general and subject-specific features. Given a test trial $\mathbf{\tilde X}$ from the target subject, the subject-general feature for the $i$-th stimulus is formulated as
\begin{equation}
\label{eq:common feature}
\rho_{1,i} = \mathrm{corr}((\mathbf{\hat w}_i^{GT})^{\mathsf{T}}\mathbf{\tilde X},(\mathbf{\hat w}_i^{GS})^{\mathsf{T}}\mathbf{Y}_i^S).
\end{equation}
The subject-specific feature is extracted by \gls{trca}
\begin{equation}
\label{eq:individual feature}
\rho_{2,i} = \mathrm{corr}((\mathbf{\hat w}_i^{T_a})^{\mathsf{T}}\mathbf{\tilde X}, (\mathbf{\hat w}_i^{T_a})^{\mathsf{T}}\mathbf{\bar X}_i^{T_a}),
\end{equation}
where $\mathbf{\hat w}_i^{T_a} = \mathrm{TRCA}(\mathbf{X}_i^{T_a})$. The final feature is integrated as 
\begin{equation}
\label{eq: final feature}
    \rho_i = \mathrm{sign}(\rho_{1,i})(\rho_{1,i})^2 + \mathrm{sign}(\rho_{2,i})(\rho_{2,i})^2.
\end{equation}

Subject-general feature is the correlation between the spatially filtered test trial and the weighted source template, measuring the similarity between the test trial and shared information in the common latent space constructed by $\mathbf{\hat w}_i^{GT}$ and $\mathbf{\hat w}_i^{GS}$. Subject-specific feature is the correlation between the spatially filtered test trial and the target subject's \gls{trc}, measuring the similarity between the test trial and the target subject's individual information. Incorporating both, the final feature balances the shared information and individual information of the test trial.

\subsubsection{Filter bank analysis}

Not only is the fundamental frequency useful for the SSVEP, but the harmonic components also contribute significantly \cite{chen_2015_PNAS_FB}. To take advantage of harmonic components, the SSVEP signals are decomposed into multiple sub-band components using the $\text{M}_3$ method in \cite{chen_2015_FBcca}, and features are calculated individually for each sub-band based on (\ref{eq: final feature}). All sub-band features are then combined to enhance recognition performance. 

The index of the sub-band is denoted by $m$, and the total number is $N_m$. For the $i$-th stimulus, the combined feature across sub-bands is given by 
\begin{equation}
\label{eq:filter bank}
r_i = \sum_{m=1}^{N_m} \alpha(m)\rho_i^{(m)},
\end{equation}
where the coefficient $\alpha(m)=m^{-1.25}+0.25$.
The stimulus frequency of the test trial can be recognized by the combined feature
\begin{equation}
\label{eq:recognition}
    \hat f = \mathop{\arg\max}\limits_{f_i} r_i, \quad i = 1,2,\ldots, N_f.
\end{equation}

\subsection{\Gls{ssitrca}}  
\label{ssitrca}
To mitigate negative transfer, a subject selection strategy is developed to select a subset of source subjects $\mathcal{S}$ which are more similar to the target subject prior to transfer learning. In this preliminary step, each source subject is assigned a similarity coefficient $c_i^n$ that quantifies the correlation between the target \gls{trc} and the source \gls{trc}
\begin{equation}
\label{eq:similarity coefficent}
    c_i^n = \mathrm{corr}(\mathbf{x}_i^{T_a}, \mathbf{y}_i^{S_n}), \quad n = 1,2,\ldots,N_{src},
\end{equation}
where $\mathbf{x}_i^{T_a}=(\mathbf{\hat w}_i^{T_a})^{\mathsf{T}}\mathbf{\bar X}_i^{T_a}$.

The subject selection module is governed by the trigger parameter $\gamma$. If $ c_i^n \leq \gamma$, $\forall n \in \{1,2,\cdots,N_{src}\}$, it indicates that all source subjects exhibit relatively low similarity to the target subject. In this case, the subject selection module is disabled as we prefer to have more data rather than selecting subjects from a less similar group, thereby all source subjects are included for transfer
\begin{equation}
\label{eq:ss-disabled}
    \mathcal{S} = \{n|n \in \{1,2,\cdots,N_{src}\}\}. 
\end{equation}
Otherwise, if $\exists n\in\{1,2,\cdots,N_{src}\}$ such that $c_i^n >\gamma$, the selection mechanism is triggered. The normalized similarity value $\Tilde{c}_i^n$ is defined as
\begin{equation}
\label{eq:normalized similarity}
    \Tilde{c}_i^n = \frac{|c_i^n|}{c_i^{max}}, 
\end{equation}
where 
\begin{equation}
    c_i^{max} = \max \{|c_i^1|, |c_i^2|, \ldots, |c_i^{N_{src}}|\}.
\end{equation}
The normalized similarity $\Tilde{c}_i^n$ lies within the range $[0,1]$. As $\Tilde{c}_i^n$ approaches 1, the source subject is more closely related to the target subject. Source subjects with $\Tilde{c}_i^n$ exceeding the lower boundary $c_{lb}\in [0,1]$ are selected for transfer
\begin{equation}
\label{eq:ss-abled}
    \mathcal{S} = \{n|\Tilde{c}_i^n > c_{lb},n \in \{1,2,\cdots,N_{src}\}\}. 
\end{equation}
Integrating the subject selection module, \gls{itrca} framework is extended to \gls{ssitrca}. 
Note that the proposed subject selection is based on \gls{trc}, which has already been computed in \gls{itrca}. This design avoids redundant calculations and reduces the additional computational cost introduced by subject selection.

The \gls{ssitrca} framework can transition to \gls{itrca} and \gls{trca} by adjusting the parameter $c_{lb}$. When $c_{lb}=0$, the set of selected source subjects would be 
\begin{align}
    \mathcal{S} &=\{n|\Tilde{c}_i^n > 0,n \in \{1,2,\cdots,N_{src}\}\} \nonumber \\  
                &=\{n|n \in \{1,2,\cdots,N_{src}\}\}.
\label{eq:ss2itrca}
\end{align}
All source subjects are included, thereby \gls{ssitrca} transitions to \gls{itrca} framework. As $c_{lb}$  increases, the stricter similarity threshold is applied, resulting in fewer source subjects being selected. Since $\mathrm{max}(\Tilde{c}_i^1, \Tilde{c}_i^2, \cdots, \Tilde{c}_i^{N_{src}})=1$, we will have at least one source subject selected when $ 0\leq c_{lb} < 1$. 
When $c_{lb}=1$, the set of selected source subjects would be 
\begin{align}
    \mathcal{S} &=\{n|\Tilde{c}_i^n > 1,n \in \{1,2,\cdots,N_{src}\}\} \nonumber \\  
                &=\emptyset.
\label{eq:ss2trca}
\end{align}
Here, no source subjects are selected, meaning that only the target subject's data is used, thereby \gls{ssitrca} transitions to \gls{trca} method.

\subsection{Evaluations}
\label{evaluation}
The proposed \gls{itrca} and \gls{ssitrca} are evaluated in both public datasets (Benchmark and BETA) and the self-collected dataset, as described in Section \ref{subsec:datasets}. Accuracy and \gls{itr} are utilized to evaluate the SSVEP recognition performance. The \gls{itr} (bits/min) represents the amount of information communicated per unit time \cite{wolpaw_2002_BCI_definition} and can be calculated by
\begin{equation}
\label{eq:itr}
    \text{ITR}=\left(\log_2{N_f} + P\log_2{P}+(1-P)\log_2 \frac{1-P}{N_f-1}\right) \frac{60}{T},
\end{equation}
where $P$ is the classification accuracy and $T$ is the time in seconds for a selection which implies the data length of signals for recognizing a stimulus, including both the gaze shifting (0.5 s) and gaze time.

\begin{figure}[!t]
    \centering
    \includegraphics[width=0.9\linewidth]{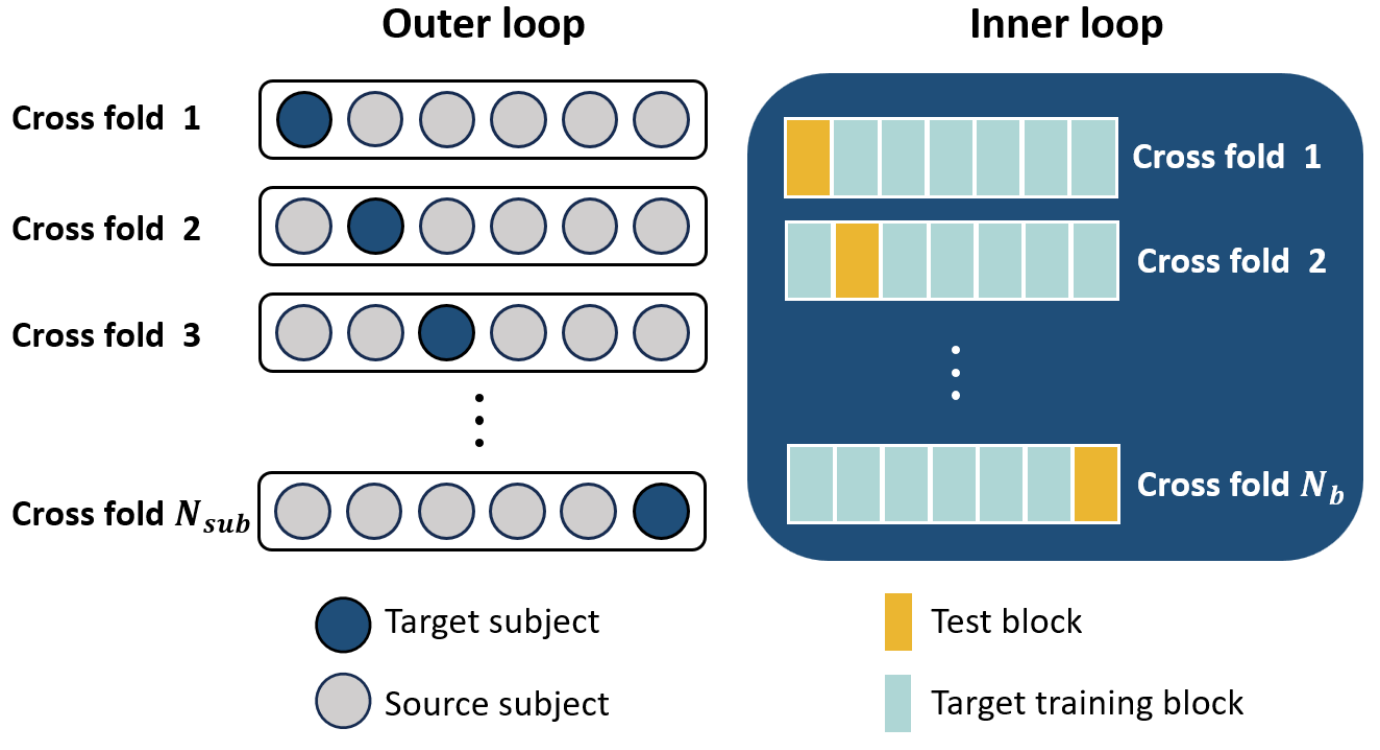}
    \caption{The illustration of leave-one-subject-out and leave-one-block-out cross-validation strategies.}
    \label{fig:cv}
\end{figure}

We utilize leave-one-out cross-validation strategies for evaluation shown in Fig. \ref{fig:cv}.  This strategy defines how the dataset is systematically partitioned for training and testing, involving two parameters inherent to the dataset: $N_{sub}$, the total number of subjects in the dataset, and $N_b$, the total number of blocks for each subject in the dataset.

In the outer loop, \gls{loso} cross-validation is applied to split the target subject and source subjects. Each subject in the dataset is iteratively treated as the target subject, while the remaining subjects serve as source subjects, i.e., $N_{src}=N_{sub}-1$. All blocks from each source subject are used for training,  i.e., $N_{sb}=N_b$. In the inner loop, \gls{lobo} cross-validation is applied for the target subject. In this step, each block (highlighted in orange) is iteratively selected as the test block, while a subset of the remaining blocks is used for training, i.e., $N_{tb}\leq N_{b}-1$. Consequently, the recognition performance of each target subject is evaluated across all blocks, and the overall performance is assessed across all subjects.

\begin{figure*}[!t]
    \centering
    \subfloat[Benchmark]{
    \label{fig:tw-benchmark}
    \includegraphics
    [scale=0.4]           
    {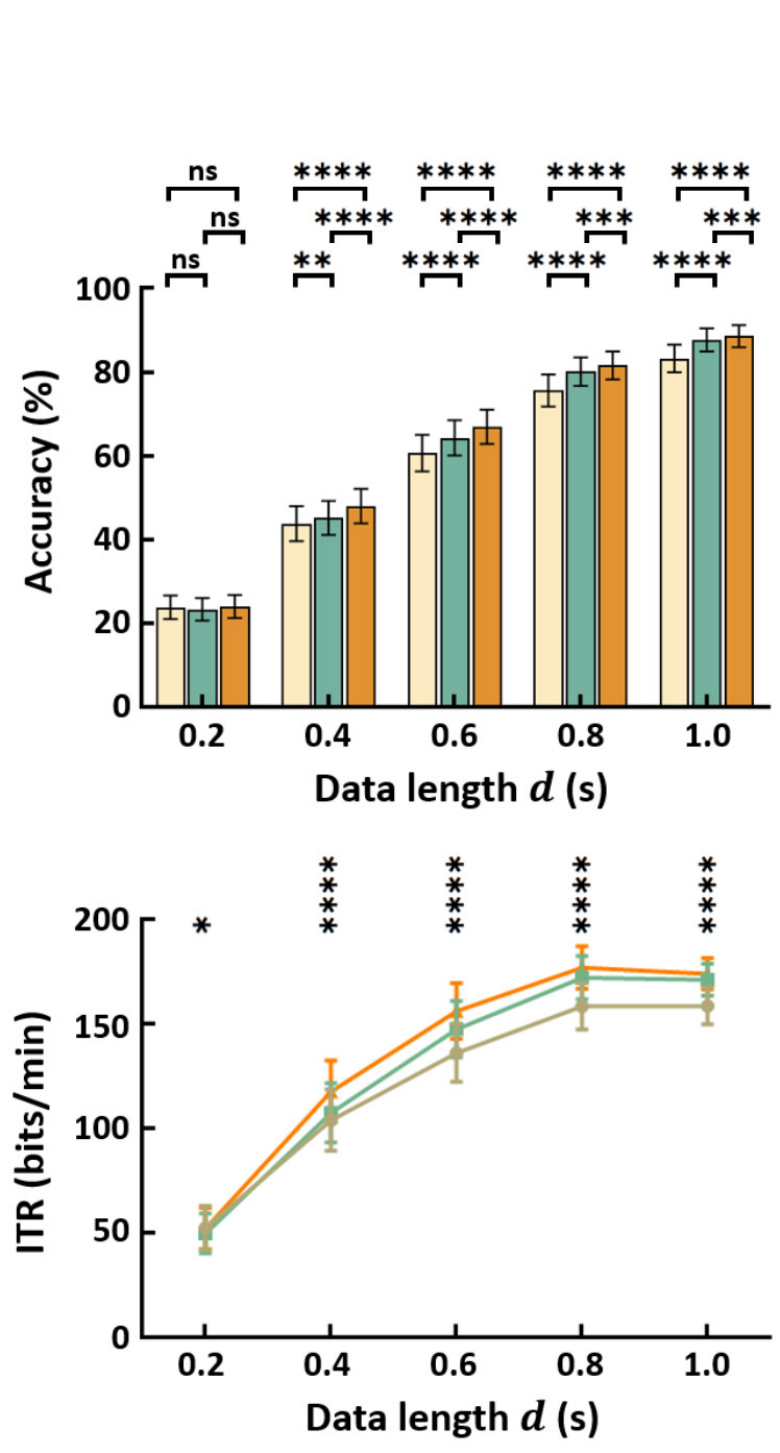}}\quad
    \subfloat[BETA]{
    \label{fig:tw-beta}
    \includegraphics
    [scale=0.4]
    {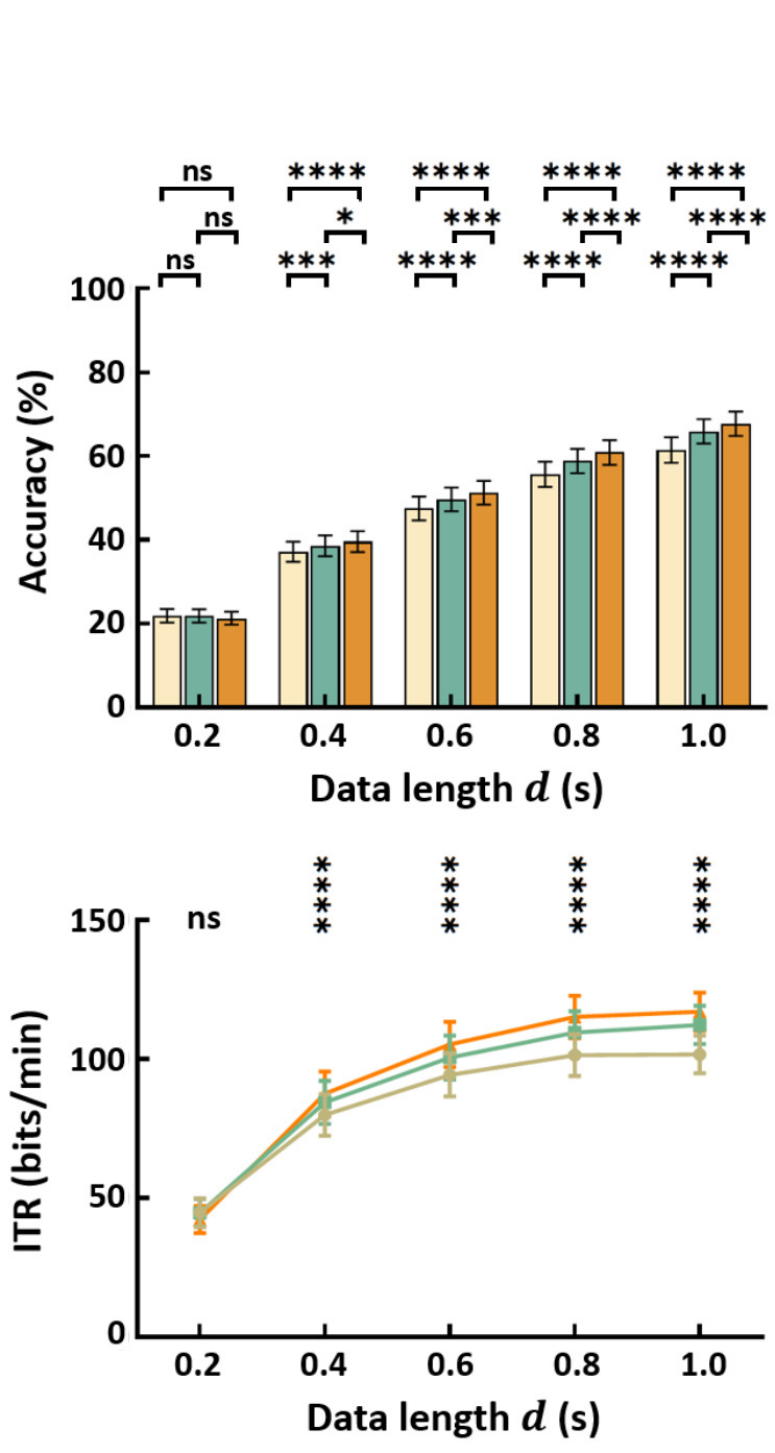}}\quad
    \subfloat[Self-collected]{
    \label{fig:tw-yue}
    \includegraphics
    [scale=0.4]
    {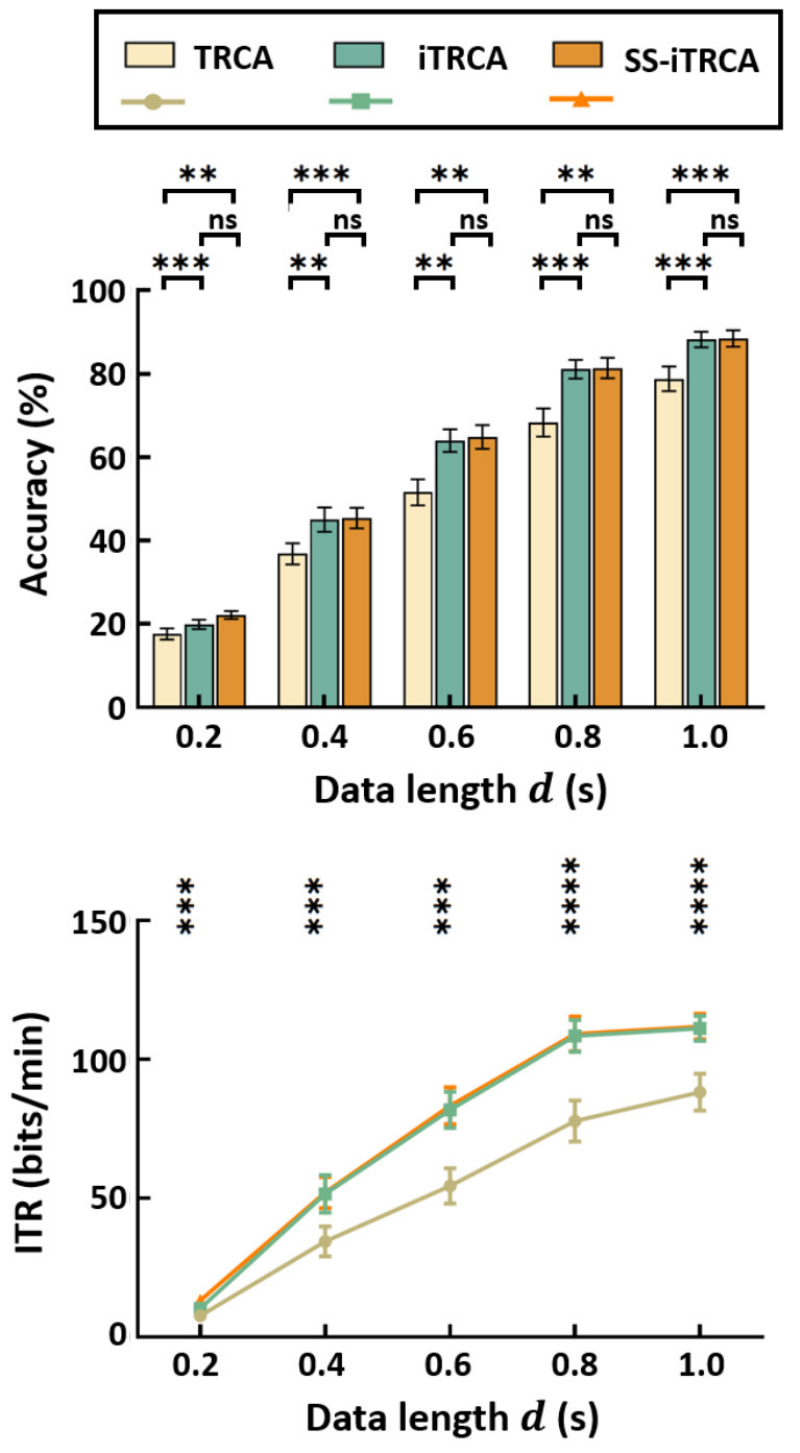}}\quad
\caption{The average recognition accuracy and ITR for TRCA, iTRCA and SS-iTRCA across all subjects for different $d$ on three datasets: (a) Benchmark, (b) BETA, and (c) Self-collected. Error bars represent \gls{sem}. Asterisks indicate the statistically significant difference between two algorithms ($*: p<0.05$, $**: p<0.01$, $***: p<0.001$, $****: p<0.0001$). Accuracy and \gls{itr} are analyzed by two-way and one-way repeated-measures ANOVA, respectively.}
\label{fig:tw}
\end{figure*}
This section evaluates the recognition performance of \gls{itrca} and \gls{ssitrca} over \gls{trca} with respect to $d$, $N_c$, and $N_{tb}$ on three datasets.

\section{Results}
\label{sec:results}

To assess transfer performance, the proposed frameworks \gls{itrca} and \gls{ssitrca} are compared to \gls{trca} method focusing on recognition accuracy, \gls{itr}, and feature distribution. Additionally, the effect of subject selection is investigated through a comparative analysis between \gls{itrca} and \gls{ssitrca}. Finally, the proposed transfer learning frameworks are examined against existing transfer learning methods \gls{transrca}, \gls{dgtf}, and \gls{cssft}. The evaluation parameters include:
\begin{itemize}
    \item data length $d$;
    \item the number of channels $N_c$;
    \item the number of target blocks $N_{tb}$;
    \item lower boundary of similarity $c_{lb}$ (for subject selection).
\end{itemize}
Each parameter is analyzed independently, with the others held constant. In this study, all available blocks in the dataset are used as source blocks, i.e., $N_{sb}=N_b$. In the filter bank analysis, the number of sub-band components $N_m$ is set to 3 and applied across all datasets.

\subsection{Recognition performance}
\label{subsec: results_recognition performance}

Fig. \ref{fig:tw} shows the average accuracy and \gls{itr} across all subjects for $d$ ranging from 0.2 s to 1 s in 0.2 s increments, with $N_c = 9$, $N_{tb} = 3$, and $c_{lb} = 0.9$. 
Across all datasets, recognition accuracy of all algorithms increases monotonically with $d$, suggesting that longer data segments provide more reliable information for recognition when $d<1$ s. 
On public datasets (Benchmark and BETA), \gls{itrca} outperforms \gls{trca} from 0.4 s onward, with \gls{ssitrca} further improving accuracy. However, the extent of improvement varies: on BETA, \gls{ssitrca} shows a more pronounced accuracy enhancement over \gls{itrca} as $d$ increases, whereas on Benchmark, the accuracy difference is less significant. On the self-collected dataset, \gls{itrca} consistently outperforms \gls{trca} across all $d$ values, while \gls{ssitrca} achieves comparable accuracy to \gls{itrca}, with no statistically significant differences observed.

Regarding \gls{itr}, the results across all datasets show a consistent pattern: \gls{itr} increases rapidly from 0.2 s to 0.6 s, peaks around 0.8 s, and either stabilizes or slightly declines at 1 s. Unlike the monotonic increase in accuracy, \gls{itr} exhibits a growth-and-decline trend as $d$ increases, reflecting the trade-off between recognition speed and accuracy: longer $d$ enhances accuracy but also reduces speed, potentially lowering overall transfer efficiency. On public datasets, SS-iTRCA achieves the highest \gls{itr} from 0.4 s onward, \gls{trca} the lowest, and \gls{itrca} lies in between. On the self-collected dataset, \gls{ssitrca} and \gls{itrca} show comparable ITR performance, while \gls{trca} consistently lags behind across all $d$ values.

\begin{figure}[t]
\centering
    \subfloat[Benchmark]{
    \label{fig:chan-benchmark}
    \includegraphics[scale=0.255]{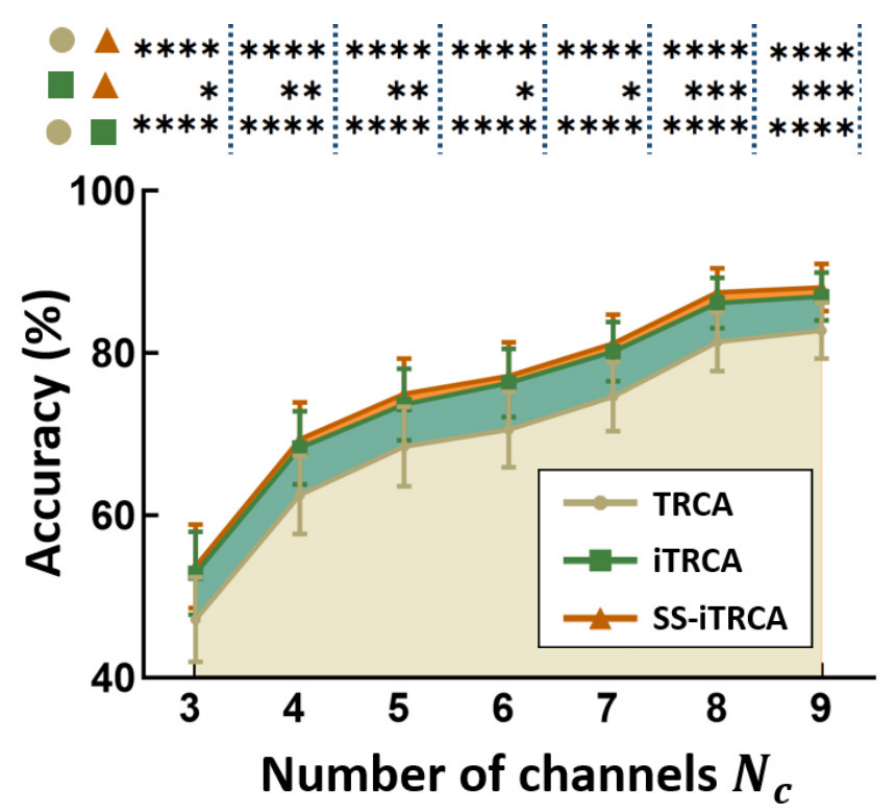}
    }
    \hspace{0.1cm}
    \subfloat[BETA]{
    \label{fig:chan-beta}
    \includegraphics[scale=0.255]{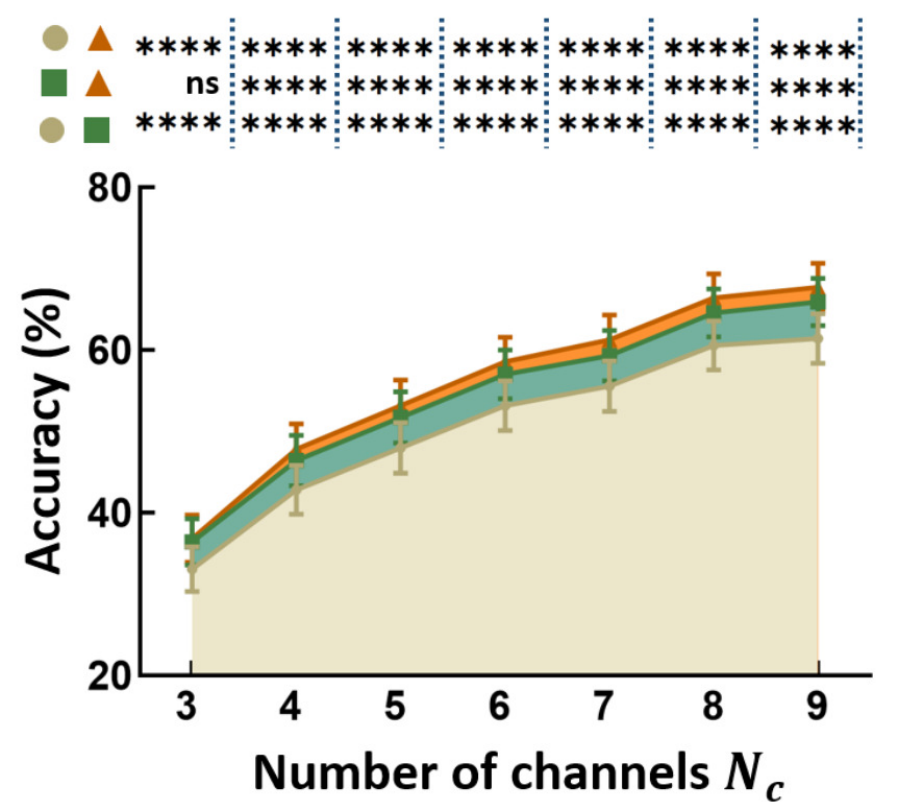}
    }   \vspace{0.3cm}\\

    \subfloat[Self-collected]{
    \label{fig:chan-yue}
    \includegraphics[scale=0.255]{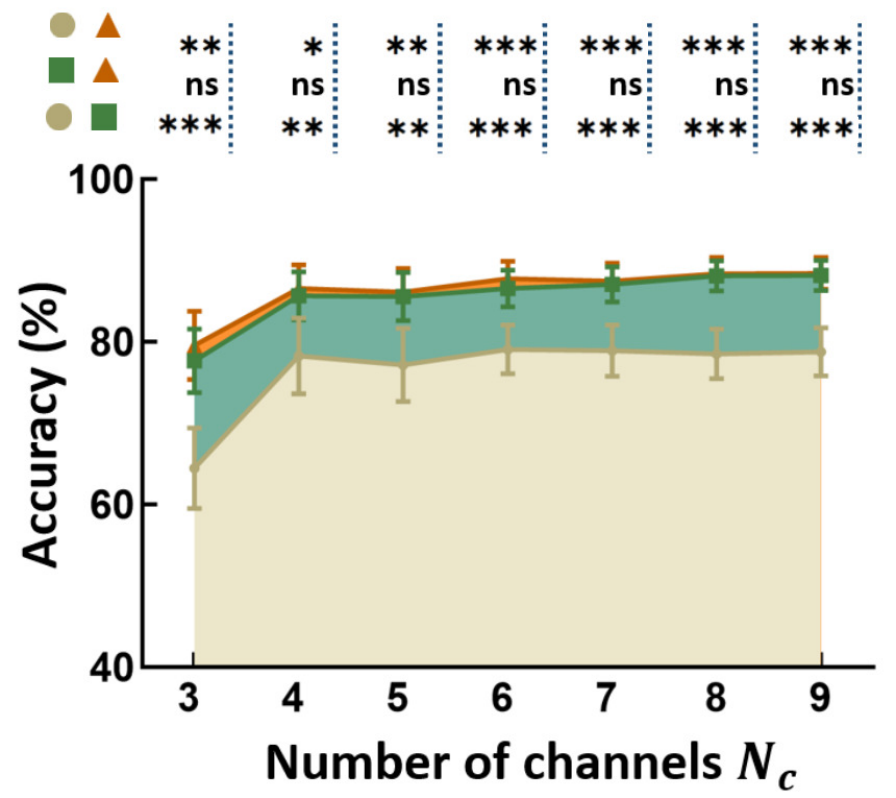}
    }
    \hspace{0.1cm}
    \subfloat[\gls{eeg} electrodes]{
    \label{fig:eeg electrodes}
    \includegraphics[scale=0.25]{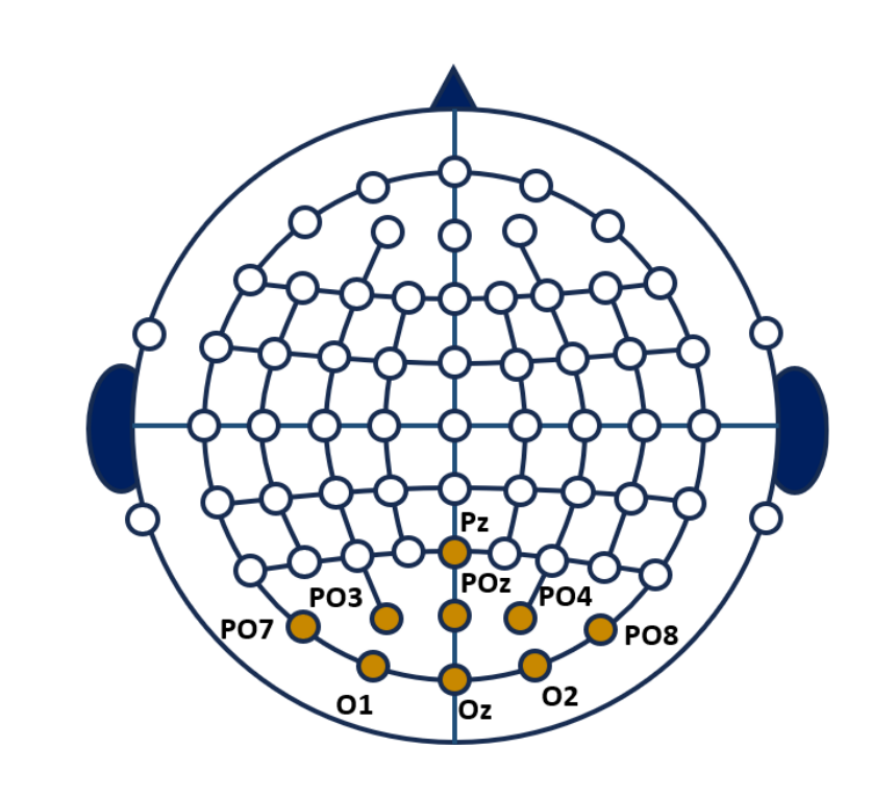}
    }
\caption{The average recognition accuracy for TRCA, iTRCA and SS-iTRCA across all subjects for different $N_c$ on three datasets: (a) Benchmark, (b) BETA, and (c) Self-collected. The nine EEG electrodes utilized in this study are highlighted in orange in (d).}

 \vspace{-0.2cm}
\label{fig:chan}
\end{figure}

\begin{table*}[!t]
    \centering
    \caption{Recognition Accuracy (\%) for Different Number of Target Blocks $N_{tb}$}
    \label{tab:Ntb}
    \begin{threeparttable}  
    \begin{tabular}{cccccccccc}
        \toprule
        \multirow{2}{*}{$\boldsymbol{N_{tb}}$}    
        &\multicolumn{3}{c}{\textbf{Benchmark}} &\multicolumn{3}{c}{\textbf{BETA}} &\multicolumn{3}{c}{\textbf{Self-collected}} \\
        \cmidrule(lr){2-4}          \cmidrule(lr){5-7}          \cmidrule(lr){8-10}
        &TRCA &SS-iTRCA &$\Delta$   &TRCA &SS-iTRCA &$\Delta$   &TRCA &SS-iTRCA &$\Delta$\\

        \midrule
        \textbf{2}   &71.06  &82.38  &11.32   
                     &49.59  &58.67  &9.08   
                     &61.52  &\textbf{82.70}  &21.19 \\
        \textbf{3}   &83.23  &\textbf{88.56}  &5.29    
                     &61.43  &67.73  &6.30   
                     &\textbf{78.75}  &\textbf{88.41}  &9.66  \\
        \textbf{4}   &\textbf{88.01}  &\textbf{91.24}  &3.22
                     &--     &--     &--     
                     &\textbf{87.42}  &90.45  &3.03  \\
        \textbf{5}   &\textbf{90.64}  &92.86  &2.21    
                     &--     &--     &--     
                     &--     &--     &--    \\
        \bottomrule
    \end{tabular}

     \begin{tablenotes}
        \item $\Delta$ is the accuracy difference between \gls{ssitrca} and \gls{trca}.
    \end{tablenotes}
    \end{threeparttable}
\end{table*}

Fig. \ref{fig:chan} presents the average accuracy across all subjects for $N_c$ ranging from 3 to 9, with $d=1$ s, $N_{tb} = 3$, and $c_{lb} = 0.9$. Nine \gls{eeg} electrodes in the occipital region (highlighted in orange in Fig. \ref{fig:eeg electrodes}) are arranged as Pz, PO7, PO3, POz, PO4, PO8, O1, Oz, and O2. The first $N_c$ electrodes are used to analyze spatial patterns; for example, when $N_c=5$, Pz, PO7, PO3, POz, and PO4 are utilized for analysis. Across all datasets, \gls{ssitrca} consistently achieves the highest accuracy, followed by \gls{itrca}, with \gls{trca} significantly lower. The difference between \gls{ssitrca} and \gls{itrca} is most pronounced on BETA and statistically significant on Benchmark, while no significant difference is observed on the self-collected dataset. Regarding the sensitivity to $N_c$, accuracy consistently improves as $N_c$ increases from 3 to 9 on Benchmark and BETA, but shows minimal improvement beyond 4 channels on the self-collected dataset.

\begin{figure}[t]
\centering
\hspace*{-0.5cm}
    \subfloat[Benchmark]{
    \label{fig:tb-benchmark}
    \includegraphics[scale=0.2]{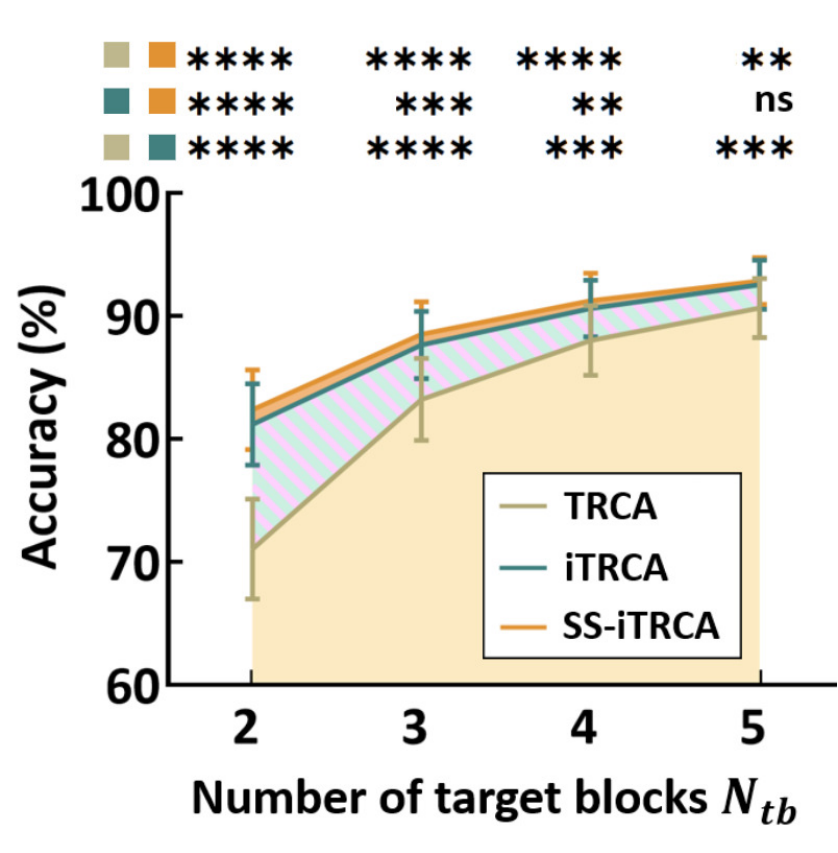}
    } 
    \subfloat[BETA]{
    \label{fig:tb-beta}
    \includegraphics[scale=0.2]{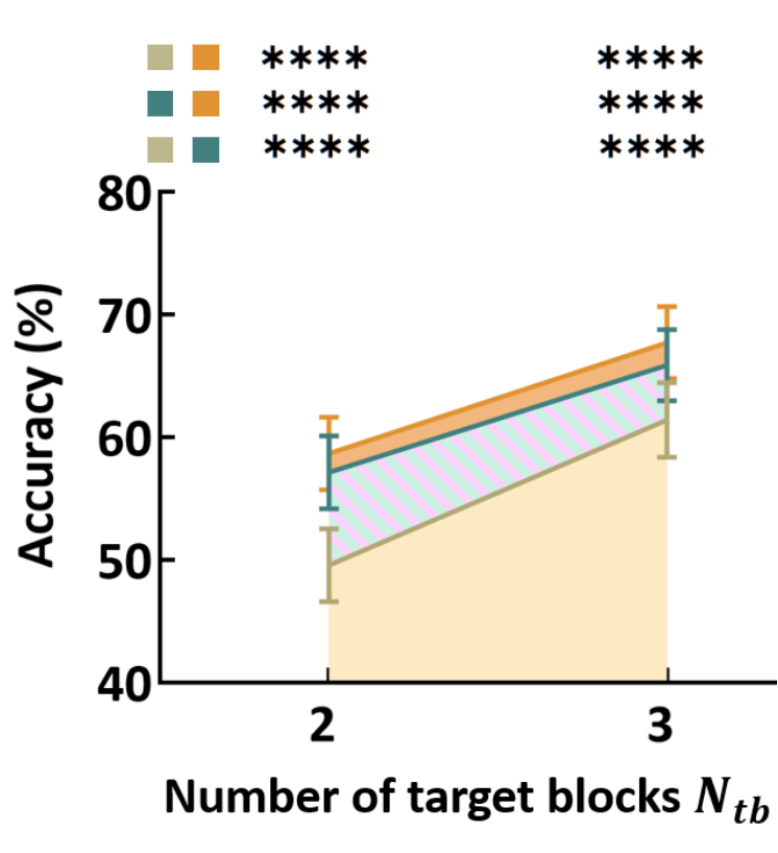}
    }
    \subfloat[Self-collected]{
    \label{fig:tb-yue}
    \includegraphics[scale=0.2]{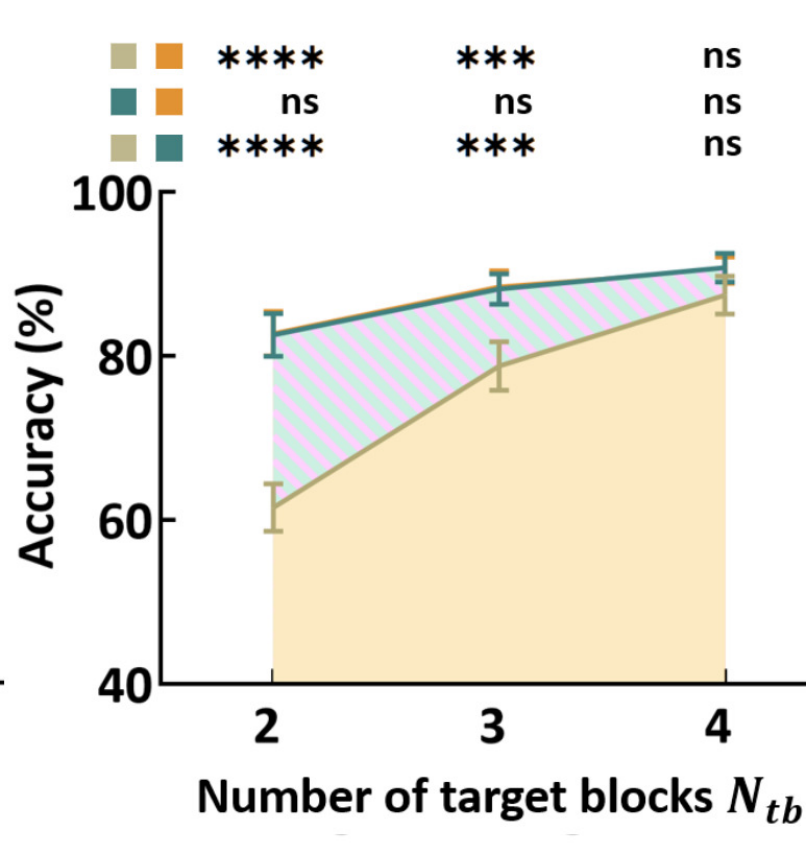}
    }
\caption{The average recognition accuracy for TRCA, iTRCA and SS-iTRCA across all subjects at different $N_{tb}$ on three datasets.}
\label{fig:tb}
\end{figure}

Fig. \ref{fig:tb} exhibits the average recognition accuracy for $N_{tb}$ ranging from 2 to $N_b-1$ ($N_b=6$, 4, and 5 in Benchmark, BETA, and self-collected dataset, respectively), with $d=1$ s, $N_c=9$, and $c_{lb}=0.9$. Across three datasets, the recognition accuracy of all algorithms increases monotonically as $N_{tb}$ increases, demonstrating that the target subject's recognition accuracy benefits from more of their own training data. On public datasets (Benchmark and BETA), \gls{ssitrca} achieves the highest accuracy, followed by \gls{itrca}, while \gls{trca} performs the worst. On the self-collected dataset, \gls{trca} remains the lowest, with \gls{itrca} and \gls{ssitrca} showing comparable accuracy and no statistically significant differences. The accuracy gap among \gls{ssitrca}, \gls{itrca}, and \gls{trca} follows a consistent trend across three datasets as $N_{tb}$ increases: the gap between \gls{ssitrca} and \gls{itrca} remains steady, while the gap between \gls{itrca} and \gls{trca} is most pronounced at $N_{tb}=2$ and gradually narrows. The difference values are organized in Table \ref{tab:Ntb}. When $N_{tb}=2$, $\Delta$ reaches 11.32\%, 9.08\%, and 21.19\% on Benchmark, BETA, and the self-collected dataset, corresponding to relative improvements of 15.93\%, 18.31\%, and 34.44\%. These results highlight the effectiveness of proposed transfer learning frameworks in scenarios with scarce target subject data.

\begin{table*}[!t]
\centering
\caption{Recognition Accuracy (\%) Comparison}
\label{tab:performance comparison}
    \begin{tabular}{cccccccccc}
    \toprule
    \multirow{2}{*}{\textbf{Parameter}} &\multicolumn{3}{c}{\textbf{Benchmark}} &\multicolumn{3}{c}{\textbf{BETA}}  &\multicolumn{3}{c}{\textbf{Self-collected}} \\
                                       \cmidrule(lr){2-4} \cmidrule(lr){5-7} \cmidrule(lr){8-10}
                                       &TRCA &iTRCA &SS-iTRCA 
                                       &TRCA &iTRCA &SS-iTRCA
                                       &TRCA &iTRCA &SS-iTRCA \\
    \midrule
    \textbf{$d$}       & 57.38 & 60.07 & 61.78    & 44.73 & 46.97 & 48.15   & 50.61 & 59.62 & 60.41 \\
    \textbf{$N_c$}     & 69.64 & 74.93 & 76.00    & 50.66 & 54.48 & 55.98   & 76.45 & 85.56 & 86.33 \\
    \textbf{$N_{tb}$}  & 83.24 & 88.01 & 88.75    & 55.51 & 61.52 & 63.20   & 75.90 & 87.16 & 87.19 \\
    \bottomrule
    \multicolumn{10}{p{251pt}}{Note: the accuracy is the average value across all cases. }
    \end{tabular}

\end{table*}

Comprehensively evaluating Figs.\ref{fig:tw}-\ref{fig:tb}, \gls{itrca} and \gls{ssitrca} consistently achieve higher accuracy across various experimental conditions ($d$, $N_c$, and $N_{tb}$), as summarized in Table \ref{tab:performance comparison}. On average, \gls{itrca} and \gls{ssitrca} improve accuracy by 4.65\% and 5.87\% over $d$, 6.07\% and 7.18\% over $N_c$, and 7.35\% and 8.17\% over $N_{tb}$, demonstrating their superior capability to leverage auxiliary data from source subjects for more accurate \gls{ssvep} recognition.

 \subsection{Feature distribution}
 \label{subsec:results_feature distribution}

\begin{figure}[!t]
\hspace*{-0.5cm}
    \subfloat[Benchmark]{
    \label{fig:tsne-benchmark}
    \includegraphics[scale=0.27]{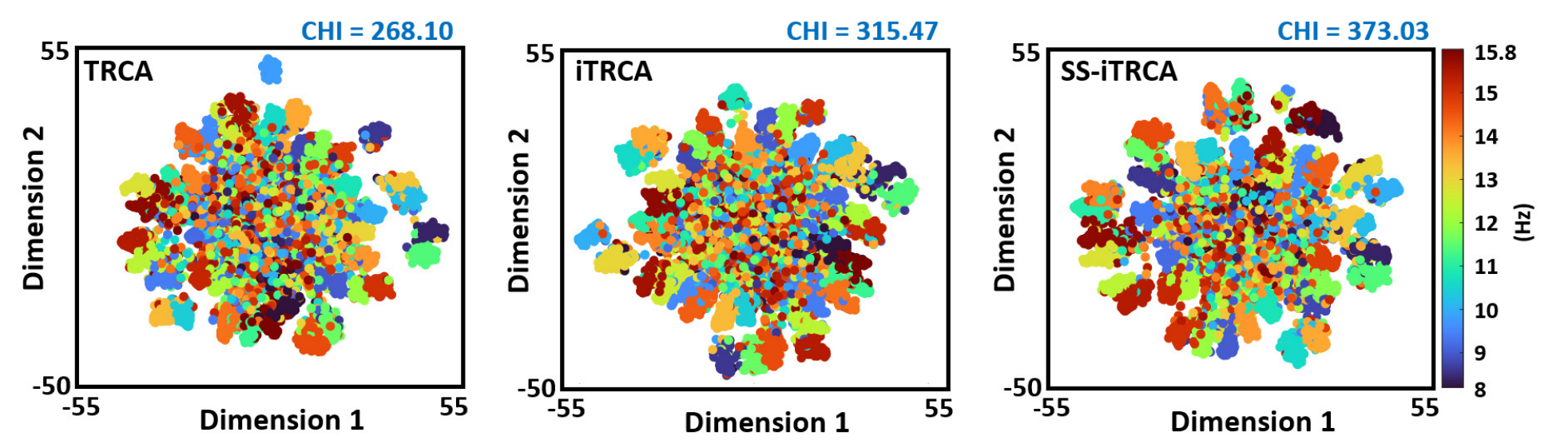}
    }   \vspace{0.1cm}\\
\hspace*{-0.5cm}
    \subfloat[BETA]{
    \label{fig:tsne-beta}
    \includegraphics[scale=0.27]{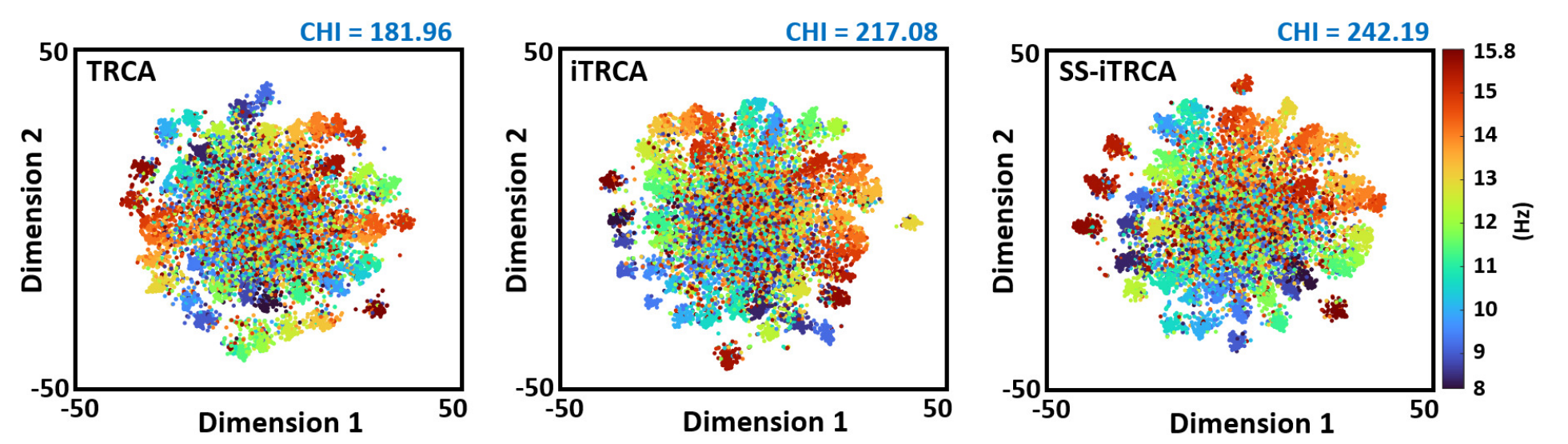}
    }    \vspace{0.1cm}\\
\hspace*{-0.5cm}
    \subfloat[Self-collected]{
    \label{fig:tsne-yue}
    \includegraphics[scale=0.27]{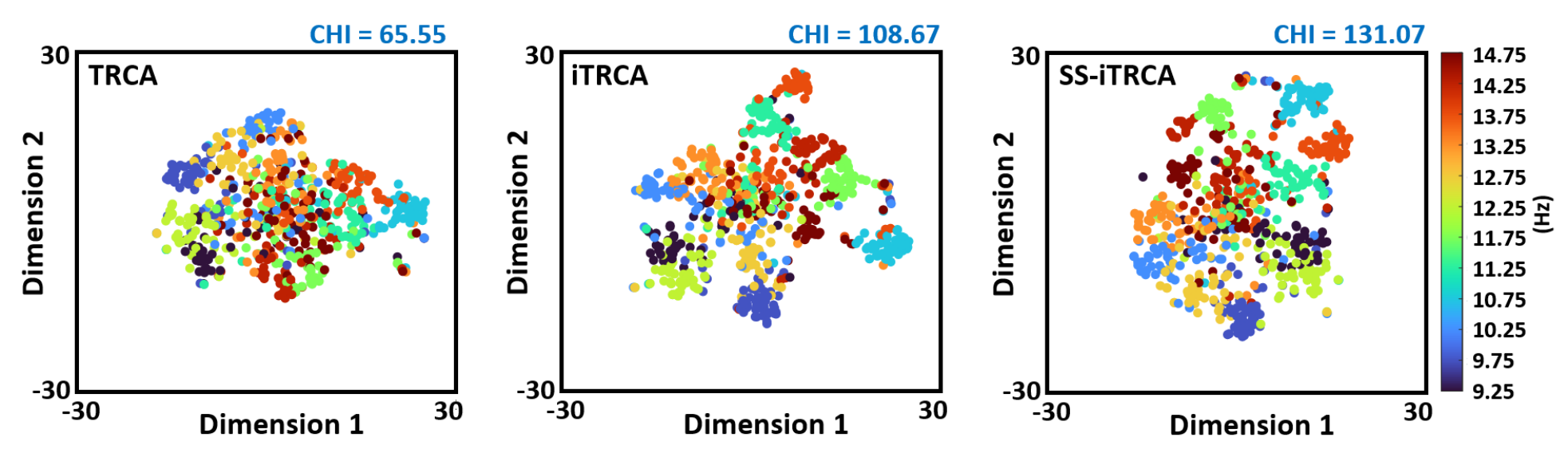}
    }
\caption{t-SNE visualization on (a) Benchmark, (b) BETA, and (c) Self-collected dataset. Each point represents a reduced 2D feature for a single trial out of $N_f \times N_b \times N_{sub}$ total trials, where $N_{sub}$ and $N_b$ refer to the iterations of \gls{loso} and \gls{lobo} cross-validation, respectively. Each stimulus includes $N_b\times N_{sub}$ trials, represented by distinct colors as shown in the color bar. \gls{chi} values are displayed in the upper-right corner of each figure to evaluate clustering performance.}
\label{fig:tsne}
\end{figure}

Fig. \ref{fig:tsne} illustrates the ability of recognition algorithm distinguishing features from different stimuli. The extracted features based on (\ref{eq:filter bank}) are reduced from high-dimensional space (40D for Benchmark and BETA, 12D for the self-collected dataset) to a 2D space using the \gls{tsne} technique \cite{van_2008_tsne}, with $d=0.6$ s, $N_c=9$, $N_{tb}=3$, and $c_{lb}=0.9$. 

The \gls{chi} \cite{calinski_1974_CHI(tsne_evaluation_metric)} is employed to assess clustering performance as the ratio of between-cluster dispersion to within-cluster dispersion, with higher \gls{chi} values indicating better clustering performance. Across all datasets, it is observed that the clusters become more compact and well-separated from the left figure (\gls{trca}) to the right (\gls{ssitrca}), reflected in higher \gls{chi} values, indicating progressive improvements in clustering quality. Comparing performance across datasets, the clusters are tighter and more distinct on Benchmark, moderately separable on BETA, and less separable on the self-collected data.

\subsection{Subject selection}
\label{subsec:results_subject selection}
The number of selected source subjects is determined by the selection trigger ($\gamma$) and similarity lower boundary ($c_{lb}$). In this study, we set the trigger $\gamma=0.5$ and examine the effect of $c_{lb}$ on recognition accuracy and the number of selected source subjects on the Benchmark dataset, with $d=0.6$ s, $N_c=9$, and $N_{tb}=3$.

\begin{figure}[!t]
\hspace*{-0.5cm}
    \subfloat[]{
    \label{fig:ss-clb}
    \includegraphics[scale=0.26]{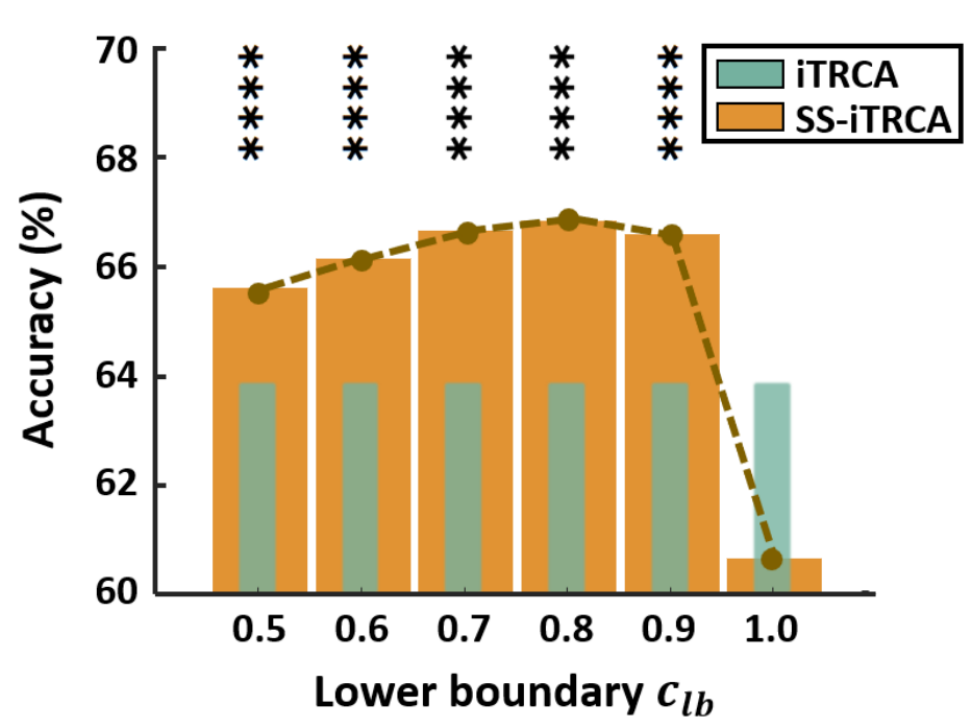}
    } 
    \subfloat[]{
    \label{fig:ss-Nsrc}
    \includegraphics[scale=0.26]{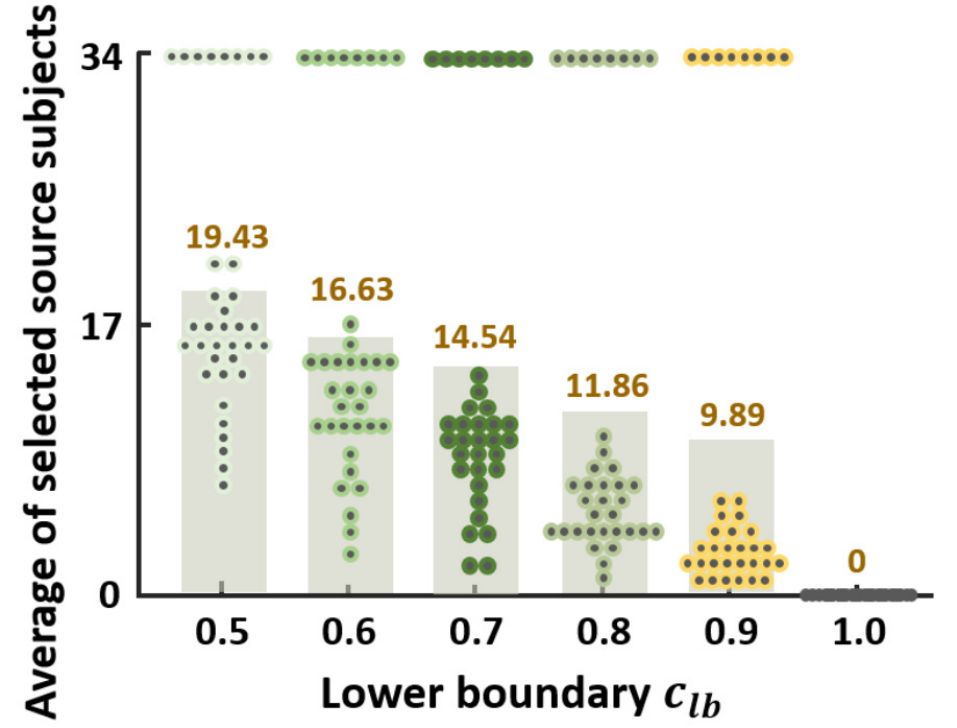}
    } \vspace{0.2cm}\\
\hspace*{-0.5cm}
    \subfloat[]{
    \label{fig:ss-allsubs}
    \includegraphics[scale=0.25]{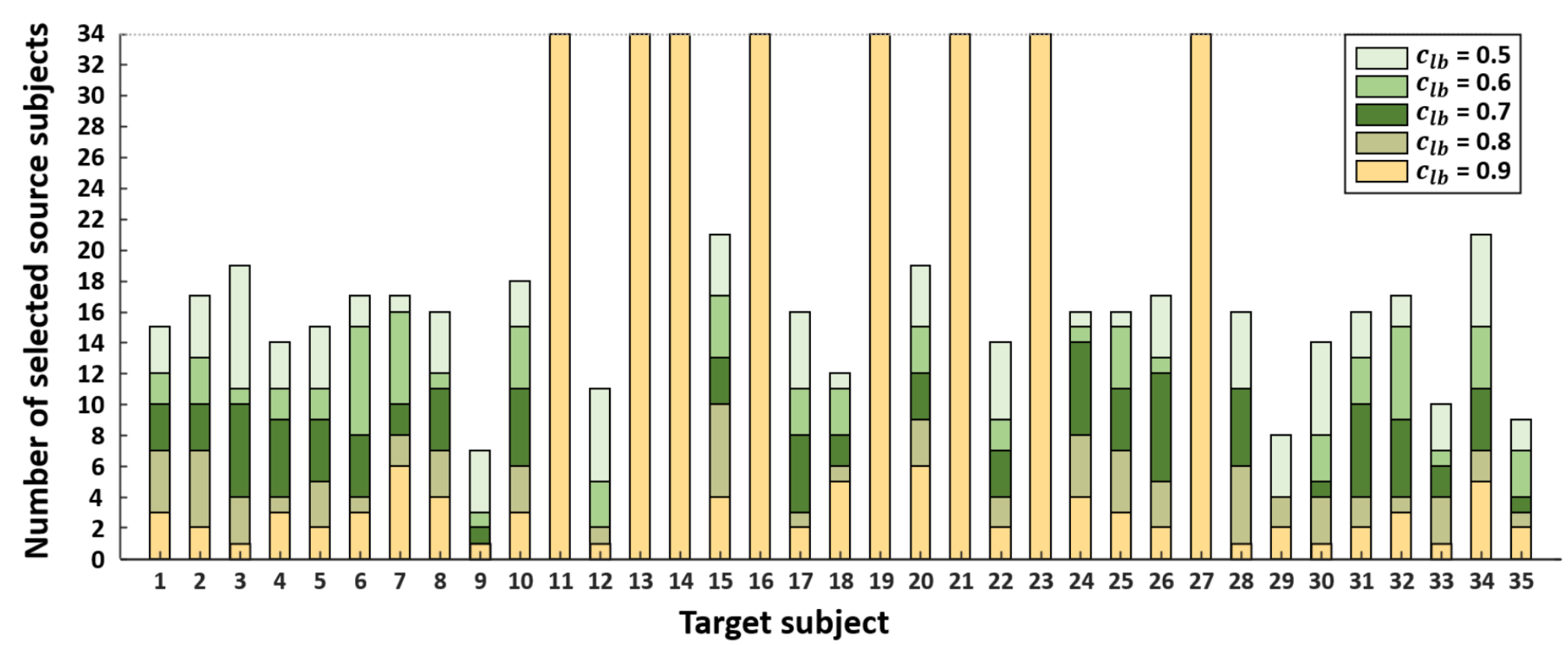}
    }
\caption{Subject selection performance on Benchmark. (a) Average recognition accuracy across all subjects. (b) The number of selected source subjects where the scatter point represents the target subject and the gray bar represents the average number across 35 target subjects. (c) The number of selected source subjects for each target subject at different thresholds $c_{lb}$. For example, for target subject 1,  the number of selected source subjects is 15, 12, 10, 7, and 3 at $c_{lb}=$ 0.5, 0.6, 0.7, 0.8, and 0.9, respectively. For target subject 11, all source subjects are included, corresponding to the case described in (16), where the subject selection strategy is not triggered.}
\label{fig:ss}
\end{figure}

Fig. \ref{fig:ss-clb} shows the average recognition accuracy across all subjects for \gls{itrca} and \gls{ssitrca}, with $c_{lb}$ ranging from 0.5 to 1. Since \gls{itrca} is unaffected by $c_{lb}$, its accuracy remains constant across all $c_{lb}$ values. In contrast, the accuracy of \gls{ssitrca} gradually increases as $c_{lb}$ rises, peaking at 0.8 before slightly declining and ultimately reducing to \gls{trca} at $c_{lb}=1$, as discussed in (\ref{eq:ss2trca}). Fig. \ref{fig:ss-Nsrc} illustrates the average number of selected source subjects as $c_{lb}$ varies, demonstrating that increasing $c_{lb}$ tightens selection criteria, leading to fewer source subjects being selected.

\subsection{Comparison with existing transfer learning method}

\begin{figure}[h]
\hspace*{-0.3cm}
    \subfloat[\centering Recognition accuracy]{
    \label{fig:TF_accuracy}
    \includegraphics
    [scale=0.3]           
    {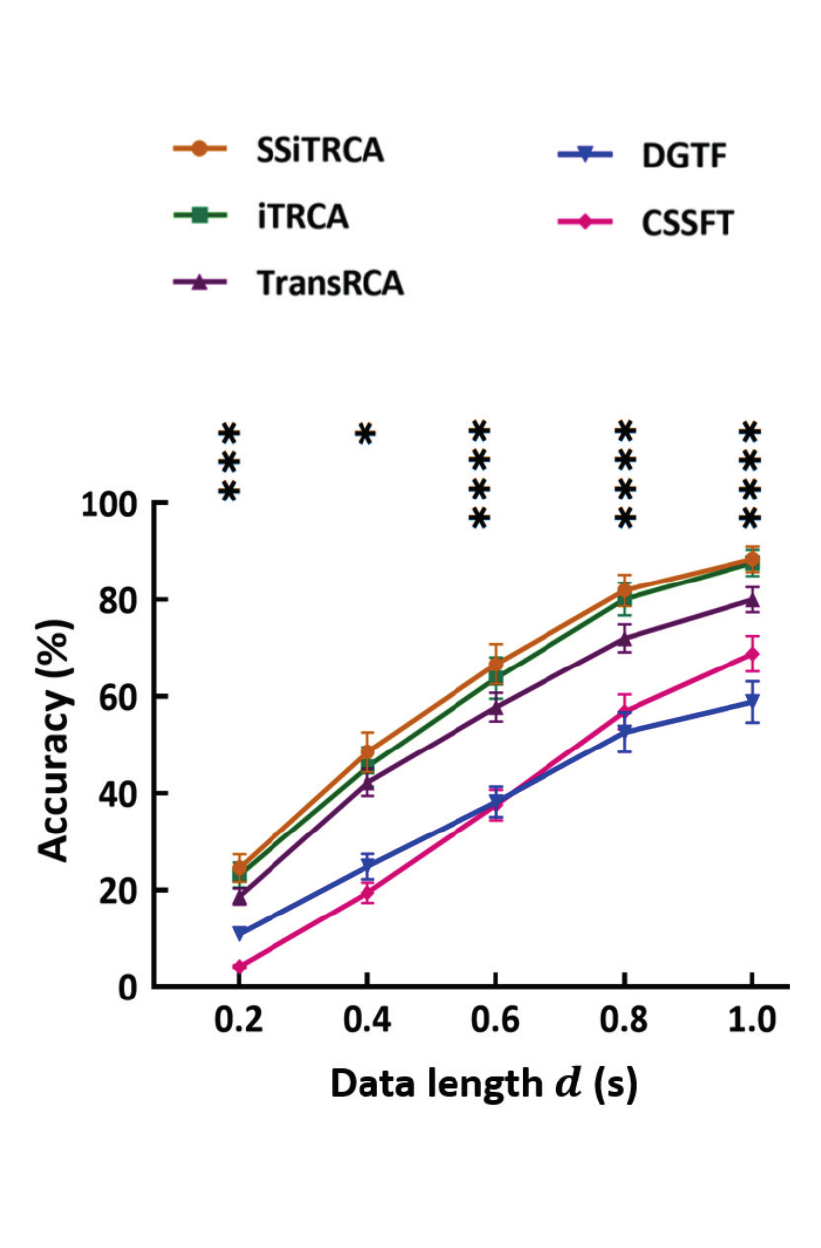}}
    \hspace{-0.1cm} 
    \subfloat[\centering Computation cost]{
    \label{fig:TF_computationCost}
    \includegraphics
    [scale=0.35]
    {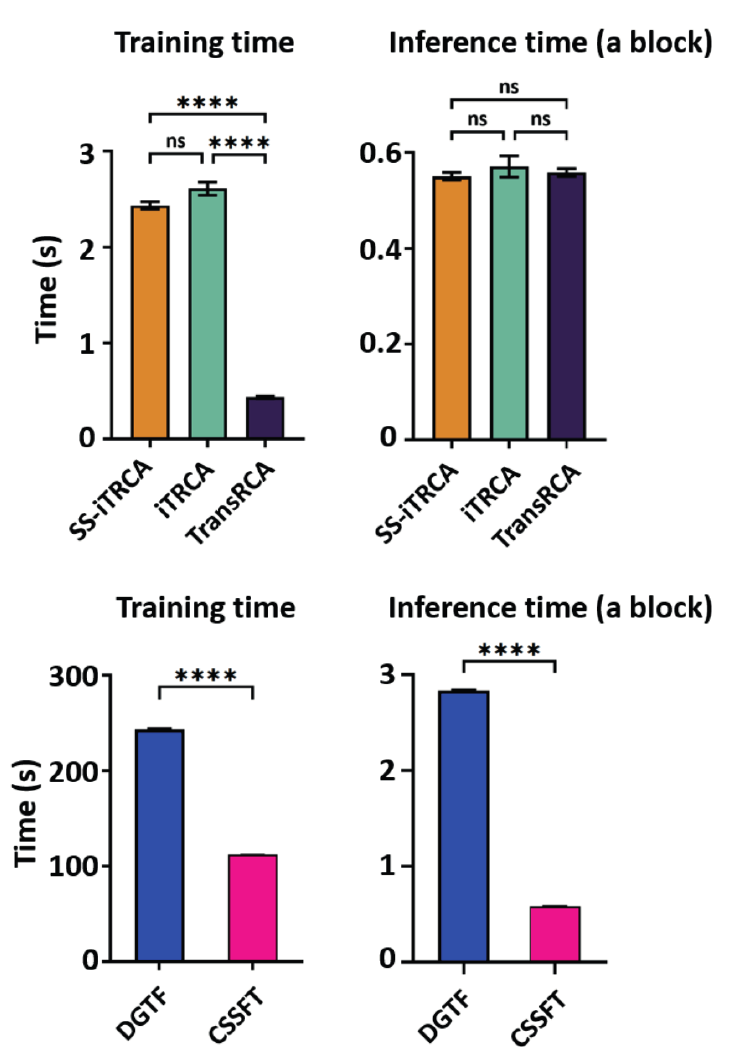}}
\caption{Performance comparison between the proposed methods (\gls{ssitrca} and \gls{itrca}) with existing transfer learning methods (\gls{transrca}, \gls{dgtf}, and \gls{cssft}). 
(a) Recognition accuracy, with a paired t-test conducted between \gls{itrca} and \gls{transrca}. (b) Training and inference time for each method, where the test time corresponds to a block comprising 40 test trials in the benchmark dataset.}
\label{fig:TF comparison}
\end{figure}

In this subsection, we further evaluate the proposed frameworks \gls{ssitrca} and \gls{itrca} against state-of-the-art transfer learning methods, \gls{transrca} \cite{lan_2023_TransRCA}, \gls{dgtf} \cite{huang_2023_domain_gen_TF}, and \gls{cssft} \cite{yan_2022_CSSFT}. Leave-one-out cross-validation strategies are employed across all methods, as described in section \ref{evaluation}. Fig. \ref{fig:TF_accuracy} presents the average recognition accuracy across all subjects for $d$ varying from 0.2 s to 1 s, with $N_c=9$, $N_{tb}=3$, and $c_{lb}=0.9$ on the benchmark dataset.
The proposed frameworks consistently outperforms \gls{transrca}, \gls{dgtf}, and \gls{cssft} across all $d$ values, with statistically significant differences.

Fig. \ref{fig:TF_computationCost} illustrates the average training and inference time with $d=$ 1 s. All experiments are conducted using MATLAB R2022a on a Windows 11 system, equipped with a 13th Intel Core i7-13700H CPU running at 2.40 GHz and an NVIDIA GeForce RTX 4070 Laptop GPU. For both training and inference stages, \gls{ssitrca}, \gls{itrca} and \gls{transrca} exhibit significantly lower computational costs compared to \gls{cssft} and \gls{dgtf}. Although \gls{transrca} achieves shorter training time than \gls{ssitrca} and \gls{itrca}, there is no statistically significant difference in inference time among them, which is crucial for real-time signal recognition. Compared with accuracy-based subject selection method \gls{cssft}, the proposed similarity-based subject selection strategy \gls{ssitrca} substantially reduces the computation cost.

Comprehensively, the proposed \gls{ssitrca} and \gls{itrca} frameworks demonstrate superior recognition performance with comparable inference time. In addition, the designed similarity-based subject selection strategy substantially enhances computational efficiency relative to accuracy-based method.

\section{Discussion}
\label{sec:discussion}

\subsection{Transfer learning performance}
\label{subsec:discuss-TF}

Section \ref{subsec: results_recognition performance} demonstrates the capability of proposed transfer learning frameworks (iTRCA and SS-iTRCA) to leverage auxiliary data from the source subjects to improve the SSVEP recognition, over the conventional TRCA. Internally, these improvements stem from the quality of extracted features, as demonstrated in Section \ref{subsec:results_feature distribution}. Features extracted by \gls{itrca} and \gls{ssitrca} exhibit a more compact distribution for the same stimulus across subjects and greater separation for different stimuli, with higher \gls{chi} values than \gls{trca}. This indicates the enhanced ability of \gls{itrca} and \gls{ssitrca} to handle inter-subject variability, a common challenge in SSVEP-based BCI systems.

The advantages of our transfer learning framework are more pronounced with limited training data from the target subject. As indicated by bold values in Table \ref{tab:Ntb}, \gls{ssitrca} reaches higher accuracy than \gls{trca} with one fewer training block, suggesting its promise as a solution for minimizing training data needs without compromising recognition accuracy.

Furthermore, the robustness of the \gls{itrca} and \gls{ssitrca} is validated across three distinct cross-subject scenarios represented by different datasets. The Benchmark dataset (35 subjects, aged 17--34) serves as a standard for assessing cross-subject generalization performance. The BETA dataset (70 subjects, aged 9--64) captures a broad range of age groups and cognitive variability, which is essential for evaluating robustness across a heterogeneous population. The self-collected dataset (11 subjects) offers a smaller and more controlled setup to test adaptability in limited data scenarios. Across all datasets, \gls{itrca} and \gls{ssitrca} consistently outperform \gls{trca}, demonstrating their versatility and effectiveness in handling diverse cross-subject scenarios.

\subsection{Subject selection-based transfer learning}

    \gls{ssitrca} is the enhanced framework of \gls{itrca} with similarity-based subject selection.
    Figs. \ref{fig:tw}-\ref{fig:tsne} consistently shows the superior performance of \gls{ssitrca} than \gls{itrca} on public datasets (Benchmark and BETA), while their performance is comparable without statistically significant differences on the self-collected dataset. This can be attributed to the small sample size (11 subjects). Under this condition, if we continue to reduce source subjects, the transfer model would struggle to extract generalized information from limited data, even if they are more similar to the target subject. This aligns with the finding from Fig. \ref{fig:ss-clb}, which suggests a balance between the similarity and the number of transferred subjects.

Additionally, Fig. \ref{fig:ss-clb} reveals the occurrence of negative transfer. As $c_{lb}$ decreases ($c_{lb}\leq0.8$), more source subjects with lower similarity to the target subject are included, leading to a decline in accuracy. This finding coincides with the negative transfer, where the dissimilar source data can hurt the target task \cite{wan_2021_TFreview} (target subject's \gls{ssvep} recognition in this study), highlighting the importance of source subject selection. Moreover, it demonstrates that the designed TRC-based metric \eqref{eq:normalized similarity} can reflect the similarity between source and target subjects, thereby improving recognition accuracy even with fewer, but more relevant, source subjects.
Fig. \ref{fig:ss-allsubs} shows how the subject selection mechanism behaves for different source subjects. It can be observed that when subjects 9 and 29 are chosen as target subjects, even when the similarity threshold is set to a low value $c_{lb}=0.5$, only a few source subjects are selected. This indicates that the SSVEP responses of these target subjects are less similar (unique) in the sense of TRC. This uniqueness may be attributed to various factors, such as subject variability, inappropriate data processing, etc.

Finally, beyond improving accuracy, \gls{ssitrca} also significantly reduces computational cost compared to the accuracy-based \gls{cssft} method (Fig. \ref{fig:TF_computationCost}), demonstrating the practical efficiency of similarity-based subject selection.

\subsection{Limitations and future directions}
While the proposed subject selection strategy performs well on public datasets (Benchmark and BETA), it does not demonstrate significant differences compared to not using it on the small-sample dataset, i.e., the self-collected dataset with 11 subjects. This limitation underscores the need for future research on effectively leveraging knowledge in source data-scarce scenarios.

In addition, deep learning-based transfer learning represents a promising future direction. However, it is important to note that the effectiveness of deep learning largely depends on the availability of large-scale training data, which still remains a challenge in EEG applications. Future studies should investigate how to adapt deep models to work efficiently under limited data conditions common in BCI scenarios.
\section{Conclusion}
\label{sec:conclusion}


In this study, we propose a transfer learning framework \gls{itrca} that accounts for the differing contributions of source subjects, and an enhanced framework \gls{ssitrca}, which incorporates a novel subject-selection strategy. \gls{itrca} and \gls{ssitrca} are evaluated on three datasets (Benchmark, BETA, and a self-collected dataset), consistently outperforming widely used \gls{trca} method in recognition accuracy and feature distribution, particularly when target subject data is limited. Additionally, \gls{ssitrca} further enhances the performance of \gls{itrca} and shows superior results compared to the existing transfer learning methods. These findings highlight \gls{itrca} and \gls{ssitrca} as promising solutions for reducing target subject data requirements without compromising accuracy \cite{chiang_2019_LST}.

\bibliographystyle{IEEEtran}
\bibliography{reference}   


\end{document}